\documentclass[peerreview]{IEEEtran}	

\usepackage{amsmath}
\usepackage{amsfonts}
\usepackage{amscd}
\usepackage{latexsym}
\usepackage{pstricks}
\usepackage{pst-node}

\usepackage[noadjust]{cite}
\usepackage{graphicx}

\usepackage{url}

\newcommand{\eref}[1]{(\ref{#1})}
\newcommand{\dist}[2]{d(#1,#2)}
\newcommand{\quo}[1]{\overline{#1}}			
\newcommand{\ind}{\B{1}}
\newcommand{\sgn}{\mathrm{sgn}\,}
\newcommand{\abs}[1]{\left|#1\right|}
\newcommand{\pr}{P}					
\newcommand{\nel}[1]{\abs{#1}}
\newcommand{\B}[1]{\mathbf{#1}}
\newcommand{\C}[1]{\mathcal{#1}}
\newcommand{\RN}{\mathbb{R}}
\newcommand{\NN}{\mathbb{N}}

\newcommand{\given}{\,|\,}
\newcommand{\atanh}{\mathrm{atanh}\,}			

\newtheorem{theo}{Theorem}
\newtheorem{coro}{Corollary}
\newtheorem{lemm}{Lemma}
\renewcommand{\QED}{\hfill\QEDopen}

\newcommand{\ve}{\C{N}}					
\newcommand{\fas}{\C{I}}				
\newcommand{\nve}{N}					
\newcommand{\des}{D}					
\newcommand{\de}[2]{#1\to#2}				
\newcommand{\ue}[2]{\{#1,#2\}}				
\newcommand{\nbe}[2]{N_{#1} \setminus #2}		
\newcommand{\nbv}[1]{N_{#1}}				
\newcommand{\nbve}[2]{N_{#1} \setminus #2}		
\newcommand{\nbfe}[2]{#1 \setminus #2}			
\newcommand{\del}[1]{{\partial{#1}}}			
\newcommand{\dele}[2]{{\partial{#1} \setminus #2}}	
\newcommand{\degr}[1]{\nel{\del{#1}}}			

\newcommand{\norm}[1]{\,\left\Vert {#1} \right\Vert\,}
\newcommand{\norms}[2]{\,\left\Vert {#1} \right\Vert_{#2}\,}
\newcommand{\mnorm}[3]{\,\left\Vert {#1} \right\Vert_{#2}^{#3}\,}
\newcommand{\lnorm}[3]{\norms{#1}{{#2}\to{#3}}}
\newcommand{\lmnorm}[5]{\mnorm{#1}{{#2}\to{#3}}{{#4}\to{#5}}}
\newcommand{\qnorm}[1]{\norm{\quo{#1}}}
\newcommand{\qnorms}[2]{\norms{\quo{#1}}{#2}}
\newcommand{\mqnorm}[3]{\mnorm{\quo{#1}}{#2}{#3}}
\newcommand{\lqnorm}[3]{\lnorm{\quo{#1}}{#2}{#3}}
\newcommand{\lmqnorm}[5]{\lmnorm{\quo{#1}}{#2}{#3}{#4}{#5}}

\newcommand{\RV}{x}					
\newcommand{\rv}[1]{\RV_{#1}}				

\newcommand{\dRV}{\C{X}}				
\newcommand{\drv}[1]{\dRV_{#1}}				

\newcommand{\POT}{\psi}
\newcommand{\pot}[1]{\POT^{#1}}				
\newcommand{\potrv}[1]{\pot{#1}(\rv{#1})}		
\newcommand{\pots}[2]{\pot{#1}_{#2}}			
\newcommand{\tpot}[1]{\tilde{\POT}^{#1}}		
\newcommand{\tpots}[2]{\tpot{#1}_{#2}}			

\newcommand{\Me}{\mu}					
\newcommand{\me}[2]{\Me^{\de{#1}{#2}}}			
\newcommand{\mervfv}[2]{\me{#1}{#2}(\rv{#2})}
\newcommand{\mervvf}[2]{\me{#1}{#2}(\rv{#1})}
\newcommand{\nMe}{\tilde{\Me}}				
\newcommand{\nme}[2]{\nMe^{\de{#1}{#2}}}		
\newcommand{\nmervfv}[2]{\nme{#1}{#2}(\rv{#2})}
\newcommand{\nmervvf}[2]{\nme{#1}{#2}(\rv{#1})}
\newcommand{\Mefp}{\Me_\infty}				
\newcommand{\mefp}[2]{\Mefp^{\de{#1}{#2}}}		
\newcommand{\mefprvfv}[2]{\mefp{#1}{#2}(\rv{#2})}
\newcommand{\mefprvvf}[2]{\mefp{#1}{#2}(\rv{#1})}

\newcommand{\lM}{\lambda}				
\newcommand{\nlM}{\tilde{\lM}}				
\newcommand{\lme}[2]{\lM^{\de{#1}{#2}}}
\newcommand{\nlme}[2]{\nlM^{\de{#1}{#2}}}
\newcommand{\lmes}[3]{\lme{#1}{#2}_{#3}}
\newcommand{\lmerv}[2]{\lme{#1}{#2}(\rv{#2})}
\newcommand{\nlmerv}[2]{\nlme{#1}{#2}(\rv{#2})}

\newcommand{\be}[1]{b_{#1}}				

\newcommand{\lV}{\C{V}}					
\newcommand{\lVl}[2]{\lV^{\de{#1}{#2}}}			
\newcommand{\lW}{\C{W}}					
\newcommand{\lWl}[2]{\lW^{\de{#1}{#2}}}			

\newcommand{\CF}{h}					
\newcommand{\tCF}{\tilde h}				
\newcommand{\cf}[2]{\CF^{#1\setminus#2}}		
\newcommand{\cfrv}[2]{\cf{#1}{#2}(\rv{\nbfe{#1}{#2}})}	
\newcommand{\cfs}[3]{\cf{#1}{#2}_{#3}}			

\newcommand{\bJ}[2]{J_{#1#2}}				
\newcommand{\bt}[1]{\theta_{#1}}			
\newcommand{\bM}{\nu}					
\newcommand{\nbM}{\tilde{\bM}}				
\newcommand{\bme}[2]{\bM^{\de{#1}{#2}}}			
\newcommand{\nbme}[2]{\nbM^{\de{#1}{#2}}}		
\newcommand{\bV}{V}					
\newcommand{\bcf}[2]{h^{#1\setminus #2}}		
\newcommand{\bcfs}[3]{h^{#1\setminus #2}_{#3}}		
\newcommand{\bcfsm}[3]{\underline{h}^{#1\setminus #2}_{#3}}		
\newcommand{\bcfsp}[3]{\overline{h}^{#1\setminus #2}_{#3}}		
\newcommand{\BCFs}[3]{\C{H}^{#1\setminus #2}_{#3}}	

\begin{document}
\title{Sufficient conditions for convergence of the Sum-Product Algorithm}
\author{Joris~M.~Mooij and Hilbert~J.~Kappen%
\thanks{J.~M.~Mooij is with the Dept.\ of Biophysics, Inst.\ for Neuroscience,
Radboud University Nijmegen, 6525 EZ Nijmegen, The Netherlands (e-mail:
\texttt{j.mooij@science.ru.nl}).}
\thanks{H.~J.~Kappen is with the Dept.\ of Biophysics, Inst.\ for Neuroscience,
Radboud University Nijmegen, 6525 EZ Nijmegen, The Netherlands (e-mail:
\texttt{b.kappen@science.ru.nl}).}
\thanks{Parts of this work have been presented at the 21st Conference on Uncertainty in Artificial Intelligence
(UAI 2005) and published in the conference proceedings \cite{MooijKappen05}.}
}

\markboth{Technical Report, Radboud University Nijmegen}{Mooij \MakeLowercase{\textit{et al.}}: Sufficient conditions for convergence of the Sum-Product Algorithm}
\maketitle

\begin{abstract}
We derive novel conditions that guarantee convergence of the Sum-Product
algorithm (also known as Loopy Belief Propagation or simply Belief Propagation)
to a unique fixed point, irrespective of the initial messages. The
computational complexity of the conditions is polynomial in the number of
variables. In contrast with previously existing conditions, our results are
directly applicable to arbitrary factor graphs (with discrete variables) and
are shown to be valid also in the case of factors containing zeros, under some
additional conditions. We compare our bounds with existing ones, numerically
and, if possible, analytically. For binary variables with pairwise
interactions, we derive sufficient conditions that take into account local
evidence (i.e.\ single variable factors) and the type of pair interactions
(attractive or repulsive). It is shown empirically that this bound outperforms
existing bounds.
\end{abstract}

\begin{keywords}
Sum-Product Algorithm, Loopy Belief Propagation, graphical models, factor graphs, marginalization, convergence, contraction, message passing
\end{keywords}

\IEEEpeerreviewmaketitle

\section{Introduction}

\PARstart{T}{he} Sum-Product Algorithm \cite{KschischangFreyLoeliger01}, also
known as Loopy Belief Propagation, which we will henceforth abbreviate as LBP,
is a popular algorithm for approximate inference on graphical models.
Applications can be found in diverse areas such as error correcting codes
(iterative channel decoding algorithms for Turbo Codes and Low Density Parity
Check Codes \cite{McElieceMacKayCheng98}), combinatorial optimization
(satisfiability problems such as 3-SAT and graph coloring
\cite{BraunsteinZecchina04}) and computer vision (stereo matching
\cite{SunZhengShum03} and image restoration \cite{Tanaka02}). LBP can be
regarded as the most elementary one in a family of related algorithms, 
consisting of double-loop algorithms \cite{HeskesAlbersKappen03}, GBP
\cite{YedidiaFreemanWeiss05}, EP \cite{Minka01}, EC \cite{OpperWinther05},
the Max-Product Algorithm \cite{WeissFreeman01}, the Survey Propagation
Algorithm \cite{BraunsteinZecchina04, BraunsteinMezardZecchina05} and Fractional
BP \cite{WiegerinckHeskes03}.  A good understanding of LBP may therefore be
beneficial to understanding these other algorithms as well.

In practice, there are two major obstacles in the application of LBP to
concrete problems: (i) if LBP converges, it is not clear whether the results
are a good approximation of the exact marginals; (ii) LBP does not always
converge, and in these cases gives no approximations at all.  These two issues
might actually be interrelated: the ``folklore'' is that failure of LBP to
converge often indicates low quality of the Bethe approximation on which it is
based. This would mean that if one has to ``force'' LBP to converge (e.g.\ by
using damping or double-loop approaches), one may expect the results to be of
low quality. 

Although LBP is an old algorithm that has been reinvented in many fields, a
thorough theoretical understanding of the two aforementioned issues and their
relation is still lacking. Significant progress has been made in recent years
regarding the question under what conditions LBP converges (e.g.
\cite{TatikondaJordan02}, \cite{Tatikonda03},
\cite{IhlerFisherWillsky04})\footnote{After submission of this work, we came to
the attention of \cite{IhlerFisherWillsky05}, which contains improved versions
of results in \cite{IhlerFisherWillsky04}, some of which are similar or
identical to results presented here (c.f.\ Section
\ref{sec:ihler_discussion}).}, on the uniqueness of fixed points
\cite{Heskes04}, and on the accuracy of the marginals \cite{Tatikonda03}, but
the theoretical understanding is still incomplete. For the special case of a
graphical model consisting of a single loop, it has been shown that convergence
rate and accuracy are indeed related \cite{Weiss00}.

In this work, we study the question of convergence of LBP and derive new
sufficient conditions for LBP to converge to a unique fixed point. Our results
are more general and in some cases stronger than previously known sufficient
conditions. 

\section{Background}

To introduce our notation, we give a short treatment of factorizing probability
distributions, the corresponding visualizations called factor graphs, and the
LBP algorithm on factor graphs. For an excellent, extensive treatment of these
topics we refer the reader to \cite{KschischangFreyLoeliger01}.

\subsection{Graphical model}

Consider $\nve$ discrete random variables $\rv{i} \in \drv{i}$, for
$i \in \ve := \{1,2,\dots,\nve\}$; we write 
$\RV = (\rv{1},\dots,\rv{\nve}) \in \dRV := \prod_{i\in \ve} \drv{i}$. 
We are interested in the class of probability measures on $\dRV$ that can
be written as a product of \emph{factors} (also called \emph{potentials}):
  \begin{equation}\label{eq:probability_measure}
  \pr(\rv{1}, \dots, \rv{\nve}) := \frac{1}{Z} \prod_{I \in \fas} \potrv{I}.
  \end{equation}
The factors $\pot{I}$ are indexed by subsets of $\ve$, i.e.\ $\fas \subseteq \C{P}(\ve)$. If
$I\in\fas$ is the subset $I = \{i_1,\dots,i_m\} \subseteq \ve$, we write 
$\rv{I} := (\rv{i_1},\dots,\rv{i_m}) \in \prod_{i\in I} \drv{i}$.  Each factor
$\pot{I}$ is a positive function\footnote{In subsection \ref{sec:factors_zeros} 
we will loosen this assumption and allow for factors containing zeros.} 
$\pot{I} : \prod_{i\in I} \drv{i} \to
(0,\infty)$. $Z$ is a normalizing constant ensuring that
$\sum_{\RV\in\dRV} \pr(\RV) = 1$. The class of probability measures described 
by \eref{eq:probability_measure} contains Markov Random Fields as well as 
Bayesian Networks.
We will use uppercase letters for indices of factors ($I,J,K,\ldots \in \fas$)
and lowercase letters for indices of variables ($i,j,k,\ldots \in \ve$). 

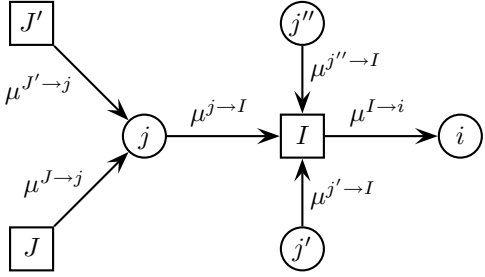
\begin{figure}[bt]
\centering
\psset{unit=1.0cm}
\begin{pspicture}(-2.0,-2.0)(4.7,2.0)
{
\psset{arrowscale=2.0}
\psframe(-1.8,-1.8)(-1.2,-1.2)
\rput[B](-1.5,-1.6){$J$}
\psframe(-1.8,1.8)(-1.2,1.2)
\rput[B](-1.5,1.4){$J'$}
\psline{->}(-1.2,-1.2)(-0.21,-0.21)
\rput[B](-1.2,-0.75){$\me{J}{j}$}
\psline{->}(-1.2,1.2)(-0.21,0.21)
\rput[B](-1.4,0.5){$\me{J'}{j}$}
\pscircle(0.0,0){0.3}
\rput[B](0.0,-0.1){$j$}
\psline{->}(0.3,0.0)(1.8,0.0)
\rput[B](1.0,0.2){$\me{j}{I}$}
\psframe(1.8,0.3)(2.4,-0.3)
\rput[B](2.1,-0.1){$I$}
\psline{->}(2.4,0.0)(3.9,0.0)
\rput[B](3.1,0.2){$\me{I}{i}$}
\pscircle(4.2,0.0){0.3}
\rput[B](4.2,-0.1){$i$}
\pscircle(2.1,-1.5){0.3}
\rput[B](2.1,-1.6){$j'$}
\psline{<-}(2.1,-0.3)(2.1,-1.2)
\rput[lB](2.2,-0.9){$\me{j'}{I}$}
\pscircle(2.1,1.5){0.3}
\rput[B](2.1,1.4){$j''$}
\psline{<-}(2.1,0.3)(2.1,1.2)
\rput[lB](2.2,0.8){$\me{j''}{I}$}
}
\end{pspicture}
\caption{\label{fig:factor_graph}
Part of the factor graph illustrating the LBP update rules
\eref{eq:mevf_update} and \eref{eq:mefv_update}. The factor nodes $I,J,J' \in
\fas$ are drawn as rectangles, the variable nodes $i,j,j',j'' \in \ve$ as
circles. Note that $\nbve{j}{I} = \{J,J'\}$ and $\nbfe{I}{i} = \{j,j',j''\}$.
Apart from the messages that have been drawn, each edge also carries a
message flowing in the opposite direction.
}
\end{figure}

The \emph{factor graph} that corresponds to the probability distribution 
\eref{eq:probability_measure} is a bipartite graph with vertex set
$\ve \cup \fas$. In the factor graph (see also Fig.\ \ref{fig:factor_graph}),
each \emph{variable node} $i \in \ve$
is connected with all the factors $I \in \fas$ that contain the variable, i.e.\
the neighbors of $i$ are the factor nodes 
$\nbv{i} := \{I \in \fas : i \in I\}$. Similarly, each \emph{factor node}
$I \in \fas$ is connected with all the variable nodes $i \in \ve$ that it
contains and we will simply denote the neighbors of $I$ by $I = \{i \in \ve:
i \in I\}$. For each variable node $i \in \ve$, we define the set of its
neighboring variable nodes by $\del{i} := \big(\bigcup \nbv{i}\big) \setminus \{i\}$, 
i.e.\ $\del{i}$ is the set of indices of those variables that interact directly
with $\rv{i}$.

\subsection{Loopy Belief Propagation}

Loopy Belief Propagation is an algorithm that calculates approximations to the
marginals $\{\pr(\rv{I})\}_{I \in \fas}$ and $\{\pr(\rv{i})\}_{i \in \ve}$ of
the probability measure \eref{eq:probability_measure}. The
calculation is done by message-passing on the factor graph: each node passes
messages to its neighbors. One usually discriminates between two types of
messages: messages $\mervfv{I}{i}$ from factors to variables and messages
$\mervvf{i}{I}$ from variables to factors (where $i \in I \in \fas$). Both
messages are positive functions on $\drv{i}$, or, equivalently, vectors in
$\RN^{\drv{i}}$ (with positive components). The messages that are sent by a
node depend on the incoming messages; the new messages, designated by $\nMe$,
are given in terms of the incoming messages by the following \emph{LBP update
rules}\footnote{ We abuse notation slightly by writing $X \setminus x$ instead
of $X \setminus \{x\}$ for sets $X$.  }
  \begin{align}
  \nmervvf{j}{I} & \propto \prod_{J \in \nbve{j}{I}} \mervfv{J}{j} \label{eq:mevf_update} \\
  \nmervfv{I}{i} & \propto \sum_{\rv{\nbfe{I}{i}}} \potrv{I} \prod_{j\in\nbfe{I}{i}} \mervvf{j}{I}.\label{eq:mefv_update}
  \end{align}
Usually, one normalizes the messages in the $\ell_1$-sense (i.e.\ such that
$\sum_{\rv{i}\in\drv{i}} \Me(\rv{i}) = 1$). If all messages have converged 
to some fixed point $\Mefp$, one
calculates the approximate marginals or \emph{beliefs}
  \begin{align*}
  \be{i}(\rv{i}) & = C^i \prod_{I \in \nbv{i}} \mefprvfv{I}{i} \approx \pr(\rv{i}) \\
  \be{I}(\rv{I}) & = C^I \potrv{I} \prod_{i \in I} \mefprvvf{i}{I} \approx \pr(\rv{I}),
  \end{align*}
where the $C^i$'s and $C^I$'s are normalization constants, chosen such that the
approximate marginals are normalized in $\ell_1$-sense.
A fixed point always exists if all factors are strictly positive
\cite{YedidiaFreemanWeiss05}. However, the existence of a fixed point does not
necessarily imply convergence towards the fixed point, and fixed points may
be unstable.

Note that the beliefs are invariant under rescaling of the messages
  \begin{equation*}
  \mefprvfv{I}{i} \mapsto \alpha^{\de{I}{i}} \mefprvfv{I}{i},\qquad \mefprvvf{i}{I} \mapsto \alpha^{\de{i}{I}} \mefprvvf{i}{I}
  \end{equation*}
for positive constants $\alpha$, which shows that the precise 
way of normalization in \eref{eq:mevf_update} and \eref{eq:mefv_update} is 
irrelevant. For numerical stability however, some way of normalization (not 
necessarily in $\ell_1$-sense) is desired to ensure that the messages stay in 
some compact domain.

In the following, we will formulate everything in terms of the messages
$\mervfv{I}{i}$ from factors to variables; the update equations are then
obtained by substituting \eref{eq:mevf_update} in \eref{eq:mefv_update}:
  \begin{equation}\label{eq:me_update}
  \nmervfv{I}{i} = C^{\de{I}{i}} \sum_{\rv{\nbfe{I}{i}}} \potrv{I} \prod_{j\in\nbfe{I}{i}} \prod_{J\in\nbve{j}{I}} \mervfv{J}{j}.
  \end{equation}
with $C^{\de{I}{i}}$ such that $\sum_{\rv{i}\in\drv{i}} \nmervfv{I}{i} = 1$.  We
consider here LBP with a \emph{parallel} update scheme, which means that all
message updates \eref{eq:me_update} are done in parallel.

\section{Special case: binary variables with pairwise interactions}

In this section we investigate the simple special case of binary variables
(i.e.\ $\nel{\drv{i}} = 2$ for all $i \in \ve$), and in addition we assume that
all potentials consist of at most two variables (``pairwise interactions'').
Although this is a special case of the more general theory to be presented
later on, we start with this simple case because it illustrates most of the
underlying ideas without getting involved with the additional technicalities of
the general case. 

We will assume that all variables are $\pm 1$-valued, i.e.\
$\drv{i} = \{-1,+1\}$ for all $i \in \ve$. We take the
factor index set as $\fas = \fas_1 \cup \fas_2$ with $\fas_1 = \ve$ (the
``local evidence'') and $\fas_2 \subseteq \{\{i,j\} : i,j \in \ve, i \ne j\}$
(the ``pair-potentials''). The probability measure
\eref{eq:probability_measure} can then be written as
  \begin{equation}\label{eq:probability_measure_binary}
  \pr(\RV) = \frac{1}{Z} \exp \left( \sum_{\{i,j\} \in \fas_2} \bJ{i}{j} \rv{i} \rv{j} + \sum_{i\in \fas_1} \bt{i} \rv{i} \right)
  \end{equation}
for some choice of the parameters $\bJ{i}{j}$ (``couplings'') and $\bt{i}$
(``local fields''), with $\potrv{i} = \exp (\bt{i} \rv{i})$ for $i \in \fas_1$
and $\pot{\{i,j\}}(\rv{i},\rv{j}) = \exp (\bJ{i}{j} \rv{i} \rv{j})$ for $\{i,j\} \in \fas_2$.

Note from \eref{eq:me_update} that the messages sent from single-variable
factors $\fas_1$ to variables are constant. Thus the question whether messages
converge can be decided by studying only the messages sent from pair-potentials
$\fas_2$ to variables. We will thus ignore messages that are sent from single-variable
factors. It turns out to be advantageous to use the following ``natural''
parameterization of the messages
  \begin{equation}\label{eq:def_bme}
  \tanh \bme{i}{j} := \me{\ue{i}{j}}{j}(\rv{j} = 1) - \me{\ue{i}{j}}{j}(\rv{j}=-1),
  \end{equation}
where $\bme{i}{j} \in \RN$ is now interpreted as a message sent from variable 
$i$ to variable $j$ (instead of a message sent from the factor $\{i,j\}$ to 
variable $j$). Note that in the pairwise case,
the product over $j \in \nbfe{I}{i}$ in \eref{eq:me_update} becomes trivial.
Some additional elementary algebraic manipulations 
show that the LBP update equations \eref{eq:me_update} become particularly 
simple in this parameterization and can be written as:
  \begin{equation}\label{eq:bme_update}
  \tanh \nbme{i}{j} = \tanh (\bJ{i}{j}) \tanh \left( \bt{i} + \sum_{t \in \dele{i}{j}} \bme{t}{i} \right)
  \end{equation}
where $\del{i} = \{t \in \ve : \{i,t\} \in \fas_2\}$ are the variables
that interact with $i$ via a pair-potential. 

Defining the set of ordered pairs $\des := \{\de{i}{j} : \ue{i}{j} \in
\fas_2\}$, we see that the parallel LBP update is a mapping $f : \RN^{\des}
\to \RN^{\des}$; \eref{eq:bme_update} specifies the component 
$\big(f(\bM)\big)^{\de{i}{j}} := \nbme{i}{j}$ in terms of the components of $\bM$.
Our goal is
now to derive sufficient conditions under which the mapping $f$ is a
contraction. For this we need some elementary but powerful mathematical
theorems. 

\subsection{Normed spaces, contractions and bounds}

In this subsection we introduce some (standard) notation and remember the
reader of some elementary but important properties of vector norms, matrix
norms, contractions and the Mean Value Theorem in arbitrary normed vector
spaces, which are the main mathematical ingredients for our basic tool, Lemma
\ref{lemm:sc_general}. The reader familiar with these topics can skip this
subsection and proceed directly to Lemma \ref{lemm:sc_general} in section
\ref{sec:basic_tool}.

Let $(V,\norm{\cdot})$ be a normed finite-dimensional real vector space.
Examples of norms that will be important later on are the $\ell_1$-norm
on $\RN^N$, defined by
  \begin{equation*}
  \norms{x}{1} := \sum_{i=1}^N \abs{x_i}
  \end{equation*}
and the $\ell_\infty$-norm on $\RN^N$, defined by
  \begin{equation*}
  \norms{x}{\infty} := \max_{i\in\{1,\dots,N\}} \abs{x_i}.
  \end{equation*}

A norm on a vector space $V$ induces a metric on $V$ by the definition
$\dist{v}{w} := \norm{v-w}$. The resulting metric space is
complete.\footnote{Completeness is a topological property which we will not
further discuss, but we need this to apply Theorem
\ref{theo:contracting_mapping_principle}.}

Let $(X,d)$ be a metric space. A mapping $f : X \to X$ is called a 
\emph{contraction with respect to $d$} if there exists $0 \le K < 1$ such that 
  \begin{equation}\label{eq:contraction}
  \dist{f(x)}{f(y)} \le K \dist{x}{y} \qquad \text{for all $x,y \in X$}.
  \end{equation}
In case $d$ is induced by a norm $\norm{\cdot}$, we will call a contraction 
with respect to $d$ a $\norm{\cdot}$-contraction.
If $(X,d)$ is complete, we can apply the following celebrated theorem, due to 
Banach:
\begin{theo}[Contracting Mapping Principle]\label{theo:contracting_mapping_principle}
Let $f : X \to X$ be a contraction of a complete metric space $(X,d)$. Then $f$
has a unique fixed point $x_\infty \in X$ and for any $x \in X$, the sequence
$x, f(x), f^2(x), \dots$ obtained by iterating $f$ converges to 
$x_\infty$. The rate of convergence is at least linear, since
$\dist{f(x)}{x_\infty} \le K \dist{x}{x_\infty}$ for all $x \in X$.
\end{theo}
\begin{proof}
Can be found in many textbooks on analysis.
\end{proof}
Note that linear convergence means that the error decreases exponentially,
indeed $\dist{x_n}{x_\infty} \le C K^n$ for some $C$.

Let ($V,\norm{\cdot})$ be a normed space. The norm induces a \emph{matrix 
norm} (also called \emph{operator norm}) on linear mappings $A : V \to V$, 
defined as follows:
  \begin{equation*}
  \norm{A} := \sup_{\substack{v \in V,\\ \norm{v} \le 1}} \norm{Av}.
  \end{equation*}
The $\ell_1$-norm on $\RN^N$ induces the following matrix norm:
  \begin{equation}\label{eq:l1_mnorm}
  \norms{A}{1} = \max_{j\in\{1,\dots,N\}}\sum_{i=1}^N \abs{A_{ij}}
  \end{equation}
where $A_{ij} := (Ae_j)_i$ with $e_j$ the $j^{\mathrm{th}}$
canonical basis vector. The $\ell_\infty$-norm on $\RN^N$
induces the following matrix norm:
  \begin{equation}\label{eq:linfty_mnorm}
  \norms{A}{\infty} = \max_{i\in\{1,\dots,N\}} \sum_{j=1}^N \abs{A_{ij}}.
  \end{equation}

In the following consequence of the well-known Mean Value Theorem, the
matrix norm of the derivative (``Jacobian'') $f'(v)$ at $v \in V$ of a
differentiable mapping $f : V \to V$ is used to bound the distance of the
$f$-images of two vectors:
%
\begin{lemm}\label{lemm:mean_value_coro}
Let $(V,\norm{\cdot})$ be a normed space and $f : V \to V$ a differentiable 
mapping. Then, for $x,y \in V$:
  \begin{equation*}
  \norm{f(y) - f(x)} \le \norm{y-x} \cdot \sup_{z \in [x,y]} \norm{f'(z)}
  \end{equation*}
where we wrote $[x,y]$ for the segment
$\{\lambda x + (1-\lambda) y : \lambda \in [0,1]\}$ joining $x$ and $y$.
\end{lemm}
\begin{proof}
See \cite[Thm.\ 8.5.4]{Dieudonne69}.
\end{proof}

\subsection{The basic tool}\label{sec:basic_tool}

Combining Theorem \ref{theo:contracting_mapping_principle} and Lemma
\ref{lemm:mean_value_coro} immediately yields our basic tool:

\begin{lemm}\label{lemm:sc_general}
Let $(V,\norm{\cdot})$ be a normed space, $f : V \to V$ differentiable and
suppose that
  \begin{equation*}
  \sup_{v\in V} \norm{f'(v)} < 1.
  \end{equation*}
Then $f$ is a $\norm{\cdot}$-contraction by Lemma \ref{lemm:mean_value_coro}. 
Hence, for any $v \in V$, the sequence $v, f(v), f^2(v), \dots$ converges to a
unique fixed point $v_\infty \in V$ with a convergence rate that is at least
linear by Theorem \ref{theo:contracting_mapping_principle}.
\QED
\end{lemm}

\subsection{Sufficient conditions for LBP to be a contraction}

We apply Lemma \ref{lemm:sc_general} to the case at hand: the parallel LBP
update mapping $f : \RN^\des \to \RN^\des$, written out in components in
\eref{eq:bme_update}. Different choices of the vector norm on $\RN^\des$
will yield different sufficient conditions for whether iterating $f$ will
converge to a unique fixed point.  We will study two examples: the $\ell_1$
norm and the $\ell_\infty$ norm.

The derivative of $f$ is easily calculated from
\eref{eq:bme_update} and is given by
  \begin{equation}\label{eq:bme_deriv}
  \Big(f'(\bM)\Big)_{\de{i}{j},\de{k}{l}} 
  = \frac{\partial \nbme{i}{j}}{\partial \bme{k}{l}} 
  = A_{\de{i}{j},\de{k}{l}} B_{\de{i}{j}}(\bM)
  \end{equation}
where\footnote{For a set $X$, we define
the indicator function $\ind_X$ of $X$ by $\ind_X(x) = 1$ if $x \in X$ and
$\ind_X(x) = 0$ if $x \not\in X$.}
  \begin{align}
  B_{\de{i}{j}}(\bM) & := \frac{1 - \tanh^2 (\bt{i} + \sum_{t\in \dele{i}{j}} \bme{t}{i})}{1 - \tanh^2 (\nbme{i}{j}(\bM))} \, \sgn \bJ{i}{j} \label{eq:bme_deriv_B} \\
  A_{\de{i}{j},\de{k}{l}} & := \tanh \abs{\bJ{i}{j}} \delta_{i,l} \ind_{\dele{i}{j}}(k). \label{eq:bme_deriv_A}
  \end{align}
Note that we have absorbed all $\bM$-dependence in the factor 
$B_{\de{i}{j}}(\bM)$; the reason for this will become apparent later on.
The factor $A_{\de{i}{j},\de{k}{l}}$ is nonnegative and independent of 
$\bM$ and captures the structure of the graphical model. Note that 
$\sup_{\bM \in \bV} \abs{B_{\de{i}{j}}(\bM)} = 1$, implying that 
  \begin{equation}\label{eq:bme_deriv_bound}
  \abs{\frac{\partial \nbme{i}{j}}{\partial \bme{k}{l}}}
  \le A_{\de{i}{j},\de{k}{l}}
  \end{equation}
everywhere on $\bV$.

\subsubsection{Example: the $\ell_\infty$-norm}

The $\ell_\infty$-norm on $\RN^\des$ yields the following condition:
\begin{coro}\label{coro:sc_linfty_binary}
For binary variables with pairwise interactions: if
  \begin{equation}\label{eq:sc_linfty_binary}
   \max_{i\in \ve} \left( (\degr{i}-1) \max_{j\in\del{i}} \tanh \abs{\bJ{i}{j}} \right) < 1,
  \end{equation}
LBP is an $\ell_\infty$-contraction and converges to a unique fixed point,
irrespective of the initial messages.
\end{coro}
\begin{proof}
Using \eref{eq:linfty_mnorm}, \eref{eq:bme_deriv_A} and \eref{eq:bme_deriv_bound}:
  \begin{equation*}\begin{split}
  \norms{f'(\bM)}{\infty} 
  & =  \max_{\de{i}{j}}\sum_{\de{k}{l}} \abs{\frac{\partial \nbme{i}{j}}{\partial \bme{k}{l}}} \\
  & \le \max_{\de{i}{j}} \sum_{\de{k}{l}} \tanh \abs{\bJ{i}{j}} \delta_{il} \ind_{\dele{i}{j}}(k) \\
  & = \max_{i\in \ve} \max_{j\in \del{i}} \sum_{k\in\dele{i}{j}} \tanh \abs{\bJ{i}{j}} \\
  & = \max_{i\in \ve} \left( (\degr{i}-1) \max_{j\in\del{i}} \tanh \abs{\bJ{i}{j}} \right),
  \end{split}\end{equation*} 
and now simply apply Lemma \ref{lemm:sc_general}.
\end{proof}

\subsubsection{Another example: the $\ell_1$-norm}

Using the $\ell_1$-norm instead, we find:
\begin{coro}\label{coro:sc_l1_binary}
For binary variables with pairwise interactions:
  \begin{equation}\label{eq:sc_l1_binary}
  \max_{i\in \ve} \max_{k\in \del{i}} \sum_{j\in\dele{i}{k}} \tanh \abs{\bJ{i}{j}}  < 1,
  \end{equation}
LBP is an $\ell_1$-contraction and converges to a unique fixed point,
irrespective of the initial messages.
\end{coro}
\begin{proof}
Similar to the proof of Corollary \ref{coro:sc_linfty_binary}, now using
\eref{eq:l1_mnorm} instead of \eref{eq:linfty_mnorm}:
  \begin{equation*}\begin{split}
  \norms{f'(\bM)}{1} 
  & \le \max_{\de{k}{l}} \sum_{\de{i}{j}} \tanh \abs{\bJ{i}{j}} \delta_{il} \ind_{\dele{i}{j}}(k) \\
  & = \max_{i\in \ve} \max_{k\in \del{i}} \sum_{j\in\dele{i}{k}} \tanh \abs{\bJ{i}{j}}.
  \end{split}\end{equation*} 
\end{proof}

It is easy to see that condition \eref{eq:sc_l1_binary} is implied by
\eref{eq:sc_linfty_binary}, but not conversely; thus in this case the 
$\ell_1$-norm yields a tighter bound than the $\ell_\infty$-norm.

\subsection{Beyond norms: the spectral radius}

Instead of pursuing a search for the optimal norm, we will derive a criterion
for convergence based on the spectral radius of the matrix
\eref{eq:bme_deriv_A}. The key idea is to look at several iterations of LBP at
once. This will yield a significantly stronger condition for convergence of LBP
to a unique fixed point. 

For a square matrix $A$, we denote by $\sigma(A)$ its \emph{spectrum}, i.e.\
the set of eigenvalues of $A$. By $\rho(A)$ we denote its \emph{spectral
radius}, which is defined as $\rho(A) := \sup \abs{\sigma(A)}$, i.e.\ the
largest modulus of eigenvalues of $A$.\footnote{One should not confuse
the spectral \emph{radius} $\rho(A)$ with the spectral \emph{norm} 
$\norms{A}{2} = \sqrt{\rho(A^T A)}$ of $A$, the matrix norm induced by the $\ell_2$-norm.}

\begin{lemm}\label{lemm:pow_contract_conv}
Let $f : X \to X$ be a mapping, $d$ a metric on $X$ and suppose that $f^N$
is a $d$-contraction for some $N \in \NN$. Then $f$ has a unique fixed point
$x_\infty$ and for any $x \in X$, the sequence $x,f(x),f^2(x),\dots$ obtained
by iterating $f$ converges to $x_\infty$.
\end{lemm}
\begin{proof}
Take any $x \in X$. Consider the $N$ sequences obtained by
iterating $f^N$, starting respectively in $x$, $f(x)$, \dots, 
$f^{N-1}(x)$:
  \begin{align*}
  & x, f^N(x), f^{2N}(x), \dots \\
  & f(x), f^{N+1}(x), f^{2N+1}(x), \dots \\
  & \vdots \\
  & f^{N-1}(x), f^{2N-1}(x), f^{3N-1}(x), \dots
  \end{align*}
Each sequence converges to $x_\infty$ since $f^N$ is a $d$-contraction
with fixed point $x_\infty$. But then the sequence 
$x, f(x), f^2(x), \dots$ must converge to $x_\infty$.
\end{proof}

\begin{theo}\label{theo:almost_Jacobi_binary}
Let $f : \RN^m \to \RN^m$ be differentiable and suppose that $f'(x) = B(x) A$,
where $A$ has nonnegative entries and $B$ is diagonal with bounded entries $\abs{B_{ii}(x)} \le 1$.
If $\rho(A) < 1$ then for any $x \in \RN^m$, the sequence $x, f(x), f^2(x),\dots$ 
obtained by iterating $f$ converges to a fixed point $x_\infty$, which does
not depend on $x$.
\end{theo}
\begin{proof}
For a matrix $B$, we will denote by $\abs{B}$ the matrix with entries
$\abs{B}_{ij} = \abs{B_{ij}}$. For two matrices $B,C$ we will write $B \le C$
if $B_{ij} \le C_{ij}$ for all entries $(i,j)$. Note that if $\abs{B} \le \abs{C}$, then
$\norms{B}{1} \le \norms{C}{1}$. Also note that $\abs{BC} \le \abs{B}\abs{C}$.
Finally, if $0 \le A$ and $B \le C$, then $AB \le AC$ and $BA \le CA$.

Using these observations and the chain rule, we have for any $n=1,2,\dots$
and any $x \in \RN^m$:
  \begin{equation*}\begin{split}
  \abs{(f^n)'(x)} & = \abs{\prod_{i=1}^{n} f'\big(f^{i-1}(x)\big)} \\
  & \le \prod_{i=1}^n \Big(\abs{B\big(f^{i-1}(x)\big)} A\Big) \le A^n,
  \end{split}\end{equation*}
hence $\norms{(f^n)'(x)}{1} \le \norms{A^n}{1}$.

By the Gelfand spectral radius theorem, 
  \begin{equation*}
  \lim_{n\to\infty} \norms{A^n}{1}^{1/n} = \rho(A).
  \end{equation*}
Choose $\epsilon>0$ such that $\rho(A) + \epsilon < 1$. For some $N$,
$\norms{A^N}{1} \le (\rho(A) + \epsilon)^N < 1$. Hence for all $x \in \RN^m$,
$\norms{(f^N)'(x)}{1} < 1$. Applying Lemma \ref{lemm:sc_general}, we 
conclude that $f^N$ is a $\ell_1$-contraction. Now apply Lemma 
\ref{lemm:pow_contract_conv}.
\end{proof}

Using \eref{eq:bme_deriv}, \eref{eq:bme_deriv_B} and \eref{eq:bme_deriv_A},
this immediately yields:

\begin{coro}\label{coro:sc_spectral_binary}
For binary variables with pairwise interactions, LBP converges to a unique 
fixed point, irrespective of the initial messages, if the spectral radius of 
the $\nel{\des}\times\nel{\des}$-matrix
  \begin{equation*}
  A_{\de{i}{j},\de{k}{l}} := \tanh \abs{\bJ{i}{j}} \delta_{i,l} \ind_{\dele{i}{j}}(k)
  \end{equation*}
is strictly smaller than 1.
\QED
\end{coro}
The calculation of the spectral norm of the (sparse) matrix $A$ can be done
using standard numerical techniques in linear algebra.

Any matrix norm of $A$ is actually an upper bound on the spectral radius
$\rho(A)$, since for any eigenvalue $\lambda$ of $A$ with eigenvector $x$ we
have $\abs{\lambda} \norm{x} = \norm{\lambda x} = \norm{Ax} \le \norm{A}
\norm{x}$, hence $\rho(A) \le \norm{A}$.  This implies that no norm in Lemma
\ref{lemm:sc_general} will result in a sharper condition than Corollary
\ref{coro:sc_spectral_binary}, hence the title of this section. 

Further, for a given matrix $A$ and some $\epsilon > 0$, there exists a vector
norm $\norm{\cdot}$ such that the induced matrix norm of $A$ satisfies
$\rho(A) \le \norm{A} \le \rho(A) + \epsilon$; see \cite{Deutsch75} for
a constructive proof. Thus 
\emph{for given $A$} one can approximate $\rho(A)$ arbitrarily close by 
induced matrix norms. This immediately gives a result on 
the convergence rate of LBP (in case $\rho(A) < 1$): for any $\epsilon > 0$,
there exists a norm-induced metric such that the linear rate of contraction
of LBP with respect to that metric is bounded from above by $\rho(A) + \epsilon$.

One might think that there is a shorter proof of Corollary
\ref{coro:sc_spectral_binary}: it seems quite plausible intuitively that in
general, for a continuously differentiable $f : \RN^m \to \RN^m$, iterating $f$
will converge to a unique fixed point if $\sup_{x\in \RN^m} \rho(f'(x)) < 1$.
However, this conjecture (which has been open for a long time) has been shown
to be true in two dimensions but false in higher dimensions
\cite{VanDenEssen97}.

\subsection{Improved bound for strong local evidence}

Empirically, it is known that the presence of strong local fields
(i.e.\ single variable factors which are far from uniform) often improves the
convergence of LBP. However, our results so far are completely independent of the 
parameters $\{\bt{i}\}_{i\in\ve}$ that measure the strength of the local evidence. 
By proceeding more carefully than we have done above, the results can easily be
improved in such a way that local evidence is taken into account.

Consider the quantity $B_{\de{i}{j}}$ defined in \eref{eq:bme_deriv_B}. 
We have bounded this quantity by noting that
$\sup_{\bM \in \bV} \abs{B_{\de{i}{j}}(\bM)} = 1$. Note that for all LBP updates
(except for the first one), the argument $\bM$ 
(the incoming messages) is in $f(\bV)$, which can be considerably smaller 
than the complete vector space $\bV$. Thus, after the first LBP update, we can use
\begin{equation*}\begin{split}
\sup_{\bM \in f(\bV)} \abs{B_{\de{i}{j}}(\bM)} 
& = \sup_{\bM \in f(\bV)} \frac{1 - \tanh^2 (\bt{i} + \sum_{k\in \dele{i}{j}} \bme{k}{i})}{1 - \tanh^2 (\nbme{i}{j}(\bM))} \\
& = \sup_{\bM \in f(\bV)} \frac{1 - \tanh^2 (\bcf{i}{j})}{1 - \tanh^2(\bJ{i}{j}) \tanh^2 (\bcf{i}{j})} 
\end{split}\end{equation*}
where we used \eref{eq:bme_update} and defined the \emph{cavity field}
\begin{equation}\label{eq:def_bcf}
\bcf{i}{j}(\bM) := \bt{i} + \sum_{k \in \dele{i}{j}} \bme{k}{i}.
\end{equation}
The function $x \mapsto \frac{1 - \tanh^2 x}{1 - \tanh^2(\bJ{i}{j}) \tanh^2 x}$ is
strictly decreasing for $x \ge 0$ and symmetric around $x = 0$, thus, defining
\begin{equation}\label{eq:def_bcfs}
\bcfs{i}{j}{*} := \inf_{\bM \in f(\bV)} \abs{\bcf{i}{j}(\bM)},
\end{equation}
we obtain
\begin{equation*}
\sup_{\bM \in f(\bV)} \abs{B_{\de{i}{j}}(\bM)} = \frac{1 - \tanh^2 (\bcfs{i}{j}{*})}{1 - \tanh^2(\bJ{i}{j}) \tanh^2 (\bcfs{i}{j}{*})}.
\end{equation*}

Now, from \eref{eq:bme_update} we derive that 
\begin{equation*}
\{\bme{k}{i} : \bM \in f(\bV)\} = (-\abs{\bJ{k}{i}}, \abs{\bJ{k}{i}}),
\end{equation*}
hence
\begin{equation*}
\{\bcf{i}{j}(\bM) : \bM \in f(\bV)\} = (\bcfs{i}{j}{-},\bcfs{i}{j}{+})
\end{equation*}
where we defined
\begin{equation*}
\bcfs{i}{j}{\pm} := \bt{i} \pm \sum_{k \in \dele{i}{j}} \abs{\bJ{k}{i}}.
\end{equation*}
We conclude that $\bcfs{i}{j}{*}$ is simply the distance between 0 and the
interval $(\bcfs{i}{j}{-},\bcfs{i}{j}{+})$, i.e.\
\begin{equation*}
\bcfs{i}{j}{*} 
= \begin{cases}
\abs{\bcfs{i}{j}{+}} & \text{if $\bcfs{i}{j}{+} < 0$} \\
\bcfs{i}{j}{-}       & \text{if $\bcfs{i}{j}{-} > 0$} \\
0                    & \text{otherwise.} 
\end{cases}
\end{equation*}
Thus the element $A_{\de{i}{j},\de{k}{i}}$ (for $i\in\del{j}, k\in\dele{i}{j}$) of the matrix $A$ 
defined in Corollary \ref{coro:sc_spectral_binary} can be replaced by
\begin{multline*}
 \tanh \abs{\bJ{i}{j}} \frac{1 - \tanh^2 (\bcfs{i}{j}{*})}{1 - \tanh^2(\bJ{i}{j}) \tanh^2 (\bcfs{i}{j}{*})} \\
 = \frac{\tanh (\abs{\bJ{i}{j}} - \bcfs{i}{j}{*}) + \tanh (\abs{\bJ{i}{j}} + \bcfs{i}{j}{*})}{2},
\end{multline*}
which is generally smaller than $\tanh \abs{\bJ{i}{j}}$ and thus gives a tighter
bound.

This trick can be repeated arbitrarily often: assume that \mbox{$m \ge 0$}
LBP updates have been done already, which means that it suffices to take the supremum
of $\abs{B_{\de{i}{j}}(\bM)}$ over $\bM \in f^m(\bV)$. Define for all 
$\de{i}{j} \in \des$ and all $t = 0, 1, \dots, m$:
\begin{align}
\bcfsm{i}{j}{t} & := \inf \{\bcf{i}{j}(\bM) : \bM \in f^t(\bV) \}, \\
\bcfsp{i}{j}{t} & := \sup \{\bcf{i}{j}(\bM) : \bM \in f^t(\bV) \},
\end{align}
and define the intervals 
\begin{equation}\label{eq:BCFs}
\BCFs{i}{j}{t} := [\bcfsm{i}{j}{t}, \bcfsp{i}{j}{t}].
\end{equation}
Specifically, for $t = 0$ we have $\bcfsm{i}{j}{0} = -\infty$ and $\bcfsp{i}{j}{0} = \infty$,
which means that 
\begin{equation}\label{eq:BCFs_0}
\BCFs{i}{j}{0} = \RN.
\end{equation}
Using \eref{eq:bme_update} and \eref{eq:def_bcf}, 
we obtain the following recursive relations for the intervals (where we use
interval arithmetic defined in the obvious way):
\begin{equation}\label{eq:BCFs_update}
  \BCFs{i}{j}{t+1} = \bt{i} + \sum_{k \in \dele{i}{j}} \atanh \left( \tanh \bJ{k}{i} \tanh \BCFs{k}{i}{t} \right).
\end{equation}
Using this recursion relation, one can calculate $\BCFs{i}{j}{m}$ and define
$\bcfs{i}{j}{*}$ as the distance (in absolute value) of the interval 
$\BCFs{i}{j}{m}$ to 0:
\begin{equation}\label{eq:bcfs}
\bcfs{i}{j}{*} 
= \begin{cases}
\abs{\bcfsp{i}{j}{m}} & \text{if $\bcfsp{i}{j}{m} < 0$} \\
\bcfsm{i}{j}{m}       & \text{if $\bcfsm{i}{j}{m} > 0$} \\
0                     & \text{otherwise.} 
\end{cases}
\end{equation}
Thus by replacing the matrix $A$ in Corollary \ref{coro:sc_spectral_binary} by
\begin{equation}\label{eq:bme_deriv_A_improved}\begin{split}
& A_{\de{i}{j},\de{k}{l}} \\
& \quad = \frac{\tanh (\abs{\bJ{i}{j}} - \bcfs{i}{j}{*}) + \tanh (\abs{\bJ{i}{j}} + \bcfs{i}{j}{*})}{2}\, \delta_{i,l} \ind_{\dele{i}{j}}(k),
\end{split}\end{equation}
we obtain stronger results that improve as $m$ increases:
\begin{coro}\label{coro:sc_spectral_improved_binary}
Let $m \ge 0$. For binary variables with pairwise interactions, LBP converges to a unique 
fixed point, irrespective of the initial messages, if the spectral radius of 
the $\nel{\des}\times\nel{\des}$-matrix defined in \eref{eq:bme_deriv_A_improved}
(with $\bcfs{i}{j}{*}$ defined in equations \eref{eq:BCFs}--\eref{eq:bcfs})
is strictly smaller than 1.
\QED
\end{coro}

\section{General case}

In the general case, when the domains $\drv{i}$ are arbitrarily large (but
finite), we do not know of a natural parameterization of the messages that
automatically takes care of the invariance of the messages $\me{I}{j}$ under
scaling (like \eref{eq:def_bme} does in the binary case). Instead
of handling the scale invariance by the parameterization and using standard
norms and metrics, it seems easier to take a simple parameterization and to
change the norms and metrics in such a way that they are insensitive to the
(irrelevant) extra degrees of freedom arising from the scale invariance. This
is actually the key insight in extending the previous results beyond the binary
case: once one sees how to do this, the rest follows in a (more or less)
straightforward way.

Another important point is to reparameterize the messages:
a natural parameterization for our analysis is now in terms of logarithms of
messages $\lme{I}{i} := \log \me{I}{i}$. The LBP update equations 
\eref{eq:me_update} can be written in terms of the log-messages as:
  \begin{equation}\label{eq:lme_update}
  \nlmerv{I}{i} = \log \sum_{\rv{\nbfe{I}{i}}} \potrv{I} \cfrv{I}{i}
  \end{equation}
where we dropped the normalization and defined
  \begin{equation}\label{eq:def_cf}
  \cfrv{I}{i} := \exp \left( \sum_{j\in\nbfe{I}{i}} \sum_{J\in \nbve{j}{I}} \lmerv{J}{j} \right).
  \end{equation}

Each log-message $\lme{I}{i}$ is a vector in the vector space 
$\lVl{I}{i} := \RN^{\drv{i}}$; we will use greek letters as indices for the
components, e.g.\ $\lmes{I}{i}{\alpha} := \lme{I}{i}(\alpha)$ with 
$\alpha \in \drv{i}$. 
We will call everything that concerns individual vector spaces $\lVl{I}{i}$ 
\emph{local} and define the \emph{global} vector space $\lV$ as the direct sum 
of the local vector spaces:
  \begin{equation*}
  \lV := \bigoplus_{i \in I \in \fas} \lVl{I}{i}
  \end{equation*}
The parallel LBP update is the mapping $f : \lV \to \lV$, written out in 
components in \eref{eq:lme_update} and \eref{eq:def_cf}.

Note that the invariance of the messages $\me{I}{i}$ under scaling amounts to
invariance of the log-messages $\lme{I}{i}$ under translation. More formally, 
defining linear subspaces
  \begin{equation}\label{eq:def_lWl}
  \lWl{I}{i} := \{ \lM \in \lVl{I}{i} : \lM_\alpha = \lM_{\alpha'} \mbox{ for all } \alpha,\alpha' \in \drv{i} \}
  \end{equation}
and their direct sum
  \begin{equation*}
  \lW := \bigoplus_{i \in I \in \fas} \lWl{I}{i} \subseteq \lV,
  \end{equation*}
the invariance amounts to the observation that
  \begin{equation*}
  f(\lM + w) - f(\lM) \in \lW \qquad \mbox{for all $\lM\in \lV$, $w \in \lW$}.
  \end{equation*}
Since $\lM+w$ and $\lM$ are equivalent for our purposes, we want our measures
of distance in $\lV$ to reflect this equivalence. Therefore we will ``divide
out'' the equivalence relation and work in the quotient space $\lV/\lW$, which is 
the topic of the next subsection.

\subsection{Quotient spaces}

Let $V$ be a finite-dimensional vector space. Let $W$ be a linear subspace
of $V$. We can consider the \emph{quotient space} $V / W := \{v + W: v \in
V\}$, where $v + W := \{v + w : w \in W\}$. Defining addition and scalar
multiplication on the quotient space in the natural way, the quotient space
is again a vector space.\footnote{Indeed, we have a null vector
$0 + W$, addition ($(v_1 + W) + (v_2 + W) := (v_1 + v_2) + W$ for 
$v_1,v_2 \in V$) and scalar multiplication 
($\lambda (v+W) := (\lambda v) + W$ for $\lambda \in \RN, v \in V$).}
We will denote its elements as $\quo{v} := v + W$. Note that the projection
$\pi : V \to V/W : v \mapsto \quo{v}$ is linear.

Let $\norm{\cdot}$ be any vector norm on $V$. It induces a \emph{quotient
norm} on $V/W$, defined by 
  \begin{equation}\label{eq:def_qnorm}
  \qnorm{v} := \inf_{w\in W} \norm{v+w},
  \end{equation}
which is indeed a norm, as one easily checks. The quotient norm in turn induces
the \emph{quotient metric} $\dist{\quo{v_1}}{\quo{v_2}} := \norm{\quo{v_2} -
\quo{v_1}}$ on $V/W$. The metric space $(V/W,d)$ is complete (since any
finite-dimensional normed vector space is complete). 

Let $f : V \to V$ be a (possibly non-linear) mapping with the following
symmetry:
  \begin{equation}\label{eq:f_symmetry}
  f(v+w) - f(v) \in W \qquad \mbox{for all $v\in V$, $w \in W$}.
  \end{equation}
We can then unambiguously define the quotient mapping 
  \begin{equation*}
  \quo{f} : V/W \to V/W : \quo{v} \mapsto \quo{f(v)},
  \end{equation*}
which yields the following commutative diagram:
  \begin{equation*}
  \begin{CD}
  V           @>f>>        V\\
  @VV{\pi}V                @VV{\pi}V \\
  V/W         @>\quo{f}>>  V/W
  \end{CD}
  \qquad\qquad \pi \circ f = \quo{f} \circ \pi
\end{equation*}

For a linear mapping $A : V \to V$, condition \eref{eq:f_symmetry} amounts to
$AW \subseteq W$, i.e.\ $A$ should leave $W$ invariant; we can then
unambiguously define the quotient mapping $\quo{A} : V/W \to V/W :
\quo{v} \mapsto \quo{Av}$. 

If $f : V \to V$ is differentiable and satisfies \eref{eq:f_symmetry},
the symmetry property \eref{eq:f_symmetry} implies that 
$f'(x) W \subseteq W$, hence we can define $\quo{f'(x)} : V/W \to V/W$.
The operation of taking derivatives is compatible with projecting onto the
quotient space. Indeed, by using the chain rule and the identity 
$\pi \circ f = \quo{f} \circ \pi$, one finds that the derivative of
the induced mapping $\quo{f} : V/W \to V/W$ at $\quo{x}$ equals the induced
derivative of $f$ at $x$:
  \begin{equation}\label{eq:deriv_quo}
  \quo{f}'(\quo{x}) = \quo{f'(x)} \qquad \mbox{for all $x \in V$}.
  \end{equation}

By Lemma \ref{lemm:sc_general}, $\quo{f}$ is a contraction with respect
to the quotient norm if 
  \begin{equation*}
  \sup_{\quo{x} \in V/W} \norm{\quo{f}'(\quo{x})} < 1.
  \end{equation*}
Using \eref{eq:def_qnorm} and \eref{eq:deriv_quo}, this condition can be written
more explicitly as:
  \begin{equation*}
  \sup_{x\in V} \sup_{\substack{v \in V,\\ \norm{v}\le 1}} \inf_{w\in W} \norm{f'(x)\cdot v + w} < 1.
  \end{equation*}

\subsection{Constructing a norm on $\lV$}

Whereas in the binary case, each message $\bme{i}{j}$ was parameterized by a 
single real number, the messages are now $\nel{\drv{i}}$-dimensional vectors 
$\lme{I}{i}$ (with components $\lmes{I}{i}{\alpha}$ indexed by 
$\alpha \in \drv{i}$). In extending the $\ell_1$-norm that provided useful in
the binary case to the more general case, we have the freedom to choose the
``local'' part of the generalized $\ell_1$-norm. Here we show how to construct
such a generalization of the $\ell_1$-norm and its properties; for a more
detailed account of the construction, see Appendix \ref{app:quotients}.

The ``global'' vector space $\lV$ is the direct sum of the ``local'' subspaces
$\lVl{I}{i}$. Suppose that for each subspace $\lVl{I}{i}$, we have a local
norm $\lnorm{\cdot}{I}{i}$. A natural generalization of the $\ell_1$-norm
in the binary case is the following global norm on $\lV$:
  \begin{equation}\label{eq:logbp_l1_norm}
  \norm{\lM} := \sum_{\de{I}{i}} \lnorm{\lme{I}{i}}{I}{i}.
  \end{equation}
It is easy to check that this is indeed a norm on $\lV$.

Each subspace $\lVl{I}{i}$ has a 1-dimensional subspace $\lWl{I}{i}$ defined in
\eref{eq:def_lWl} and the local norm on $\lVl{I}{i}$ induces a local quotient
norm on the quotient space $\lVl{I}{i} / \lWl{I}{i}$. The global norm
\eref{eq:logbp_l1_norm} on $\lV$ induces a global quotient norm on $\lV/\lW$,
which is simply the sum of the local quotient norms (c.f.\ 
\eref{eq:gen_l1_qnorm}):
  \begin{equation}\label{eq:logbp_l1_qnorm}
  \qnorm{\lM} = \sum_{\de{I}{i}} \lqnorm{\lme{I}{i}}{I}{i}.
  \end{equation}

Let $\lM \in \lV$. The derivative $f'(\lM)$ of $f : \lV \to \lV$ at $\lM$
is a linear mapping $f'(\lM) : \lV \to \lV$ satisfying $f'(\lM) \lW \subseteq
\lW$. It projects down to a linear mapping $\quo{f'(\lM)} : \lV / \lW \to \lV /
\lW$. The matrix norm of $\quo{f'(\lM)}$ induced by the quotient norm
\eref{eq:logbp_l1_qnorm} is given by (c.f.\ \eref{eq:gen_l1_qmnorm}):
  \begin{equation}\label{eq:logbp_l1_qmnorm}
  \qnorm{f'(\lM)} = \max_{\de{J}{j}} \sum_{\de{I}{i}} \lmqnorm{\big(f'(\lM)\big)_{\de{I}{i},\de{J}{j}}}{I}{i}{J}{j}
  \end{equation}
where the local quotient matrix norm of the ``block'' 
$\big(f'(\lM)\big)_{\de{I}{i},\de{J}{j}}$ is given by (c.f.\ 
\eref{eq:gen_l1_lmqnorm}):
  \begin{equation}\label{eq:logbp_l1_lmqnorm}
  \begin{split}
  & \lmqnorm{\big(f'(\lM)\big)_{\de{I}{i},\de{J}{j}}}{I}{i}{J}{j} \\
   & \qquad = \sup_{\substack{v \in \lVl{J}{j},\\ \lnorm{v}{J}{j}\le 1}} \lqnorm{\big(f'(\lM)\big)_{\de{I}{i},\de{J}{j}} v}{I}{i}.
  \end{split}\end{equation}

The derivative of the (unnormalized) parallel LBP update 
\eref{eq:lme_update} is easily calculated:
  \begin{equation}\label{eq:me_deriv}\begin{split}
  & \frac{\partial \nlmerv{I}{i}}{\partial \lme{J}{j}(y_j)} = \ind_{\nbve{j}{I}}(J) \ind_{\nbfe{I}{i}}(j)  \\
  & \ \times \frac{\sum_{\rv{\nbfe{I}{i}}} \pot{I}(\rv{i},\rv{j},\rv{I\setminus \{i,j\}}) \delta_{\rv{j},y_j} \cfrv{I}{i}}{\sum_{\rv{\nbfe{I}{i}}} \pot{I}(\rv{i},\rv{I\setminus i}) \cfrv{I}{i}}.
  \end{split}\end{equation}
To lighten the notation, we will use greek subscripts
instead of arguments: let $\alpha$ correspond to $\rv{i}$, $\beta$ to $\rv{j}$,
$\beta'$ to $y_j$ and $\gamma$ to $\rv{\nbfe{I}{\{i,j\}}}$; for example, 
we write $\cfrv{I}{i}$ as $\cfs{I}{i}{\beta\gamma}$. Taking the global 
quotient norm \eref{eq:logbp_l1_qmnorm} of \eref{eq:me_deriv} yields:
  \begin{equation}\label{eq:me_deriv_qmnorm}
  \qnorm{f'(\lM)} = \max_{\de{J}{j}} \sum_{\de{I}{i}} \ind_{\nbve{j}{I}}(J) \ind_{\nbfe{I}{i}}(j) B_{\de{I}{i},\de{J}{j}}\big(\cf{I}{i}(\lM)\big)
  \end{equation}
where
  \begin{equation}\label{eq:me_deriv_B}
  B_{\de{I}{i},\de{J}{j}}\big(\cf{I}{i}(\lM)\big) := \lmqnorm{\frac{\sum_\gamma \pots{I}{\alpha\beta'\gamma} \cfs{I}{i}{\beta'\gamma}(\lM)}{\sum_\beta \sum_\gamma \pots{I}{\alpha\beta\gamma} \cfs{I}{i}{\beta\gamma}(\lM)}}{I}{i}{J}{j}.
  \end{equation}
Note that $B_{\de{I}{i},\de{J}{j}}$ depends on $\lM$ via the dependence of
$\cf{I}{i}$ on $\lM$ (c.f.\ \eref{eq:def_cf}). We will for the moment simplify 
matters by assuming that $\lM$ can be any vector in $\lV$, and later discuss
the more careful estimate (where $\lM \in f^m(\lV)$):
  \begin{equation}\label{eq:me_deriv_A}
  \sup_{\lM\in\lV} B_{\de{I}{i},\de{J}{j}}\big(\cf{I}{i}(\lM)\big) \le 
  \sup_{\cf{I}{i} > 0} B_{\de{I}{i},\de{J}{j}}(\cf{I}{i}).
  \end{equation}
Defining the matrix $A$ by the expression on the r.h.s.\ and using
\eref{eq:logbp_l1_lmqnorm} and \eref{eq:def_qnorm}, we obtain: 
  \begin{equation}\label{eq:me_deriv_A_expl}\begin{split}
  &A_{\de{I}{i},\de{J}{j}} := \sup_{\cf{I}{i} > 0} B_{\de{I}{i},\de{J}{j}}(\cf{I}{i}) = \\
  &\sup_{\cf{I}{i}>0} \sup_{\substack{v \in \lVl{J}{j} \\ \lnorm{v}{J}{j} \le 1}} \inf_{w\in \lWl{I}{i}} 
  \lnorm{\frac{\sum_{\beta'}\sum_\gamma \pots{I}{\alpha\beta'\gamma} \cfs{I}{i}{\beta'\gamma} v_{\beta'}}{\sum_\beta \sum_\gamma \pots{I}{\alpha\beta\gamma} \cfs{I}{i}{\beta\gamma}} - w}{I}{i}
  \end{split}\end{equation}
for $\de{I}{i}$ and $\de{J}{j}$ such that $j \in \nbfe{I}{i}$ and $J \in \nbve{j}{I}$.
Surprisingly, it turns out that we can calculate \eref{eq:me_deriv_A_expl}
analytically if we take all local norms to be $\ell_\infty$ norms.  We have also tried the
$\ell_2$ norm and the $\ell_1$ norm as local norms, but were unable to
calculate expression \eref{eq:me_deriv_A_expl} analytically in these cases.
Numerical calculations turned out to be difficult because of the
nested suprema.

\subsection{Local $\ell_\infty$ norms}

Take for all local norms $\lnorm{\cdot}{I}{i}$ the $\ell_\infty$ norm on 
$\lVl{I}{i} = \RN^{\drv{i}}$. The local subspace $\lWl{I}{i}$
is spanned by the vector $\ind := (1,1,\dots,1) \in \RN^{\drv{i}}$. The
local quotient norm of a vector $v \in \lVl{I}{i}$ is thus 
given by
  \begin{equation}\label{eq:linfty_qnorm}\begin{split}
  \lqnorm{v}{I}{i}
  & = \qnorms{v}{\infty} 
  = \inf_{w\in\RN}\norms{v+w\ind}{\infty} \\
  & = \frac{1}{2} \sup_{\alpha,\alpha' \in \drv{i}} \abs{v_\alpha - v_{\alpha'}}.
  \end{split}\end{equation}
For a linear mapping $A : \lVl{J}{j} \to \lVl{I}{i}$ that satisfies
$A\lWl{J}{j} \subseteq \lWl{I}{i}$, the induced quotient matrix norm
\eref{eq:logbp_l1_lmqnorm} is given by
  \begin{equation}\label{eq:linfty_lmqnorm}\begin{split}
  \lmqnorm{A}{I}{i}{J}{j} 
  & = \sup_{\substack{v \in \lVl{J}{j},\\ \norms{v}{\infty}\le 1}} \qnorms{Av}{\infty} \\
  & = \sup_{\substack{v \in \lVl{J}{j},\\ \norms{v}{\infty} \le 1}} \frac{1}{2} \sup_{\alpha,\alpha'\in\drv{i}} \bigg|\sum_\beta (A_{\alpha\beta} - A_{\alpha'\beta}) v_\beta\bigg| \\
  & = \frac{1}{2} \sup_{\alpha,\alpha'\in\drv{i}} \sum_\beta \abs{A_{\alpha\beta} - A_{\alpha'\beta}} \\
  \end{split}\end{equation}

Fixing for the moment $\de{I}{i}$ and $\de{J}{j}$ (such that $j \in
\nbfe{I}{i}$ and $J \in \nbve{j}{I}$) and dropping the superscripts from the
notation, using \eref{eq:linfty_lmqnorm}, we 
can write \eref{eq:me_deriv_A_expl} as
  \begin{equation*}
  \sup_{\CF> 0} \frac{1}{2} \sup_{\alpha,\alpha' \in \drv{i}} \sum_\beta \abs{\frac{\sum_\gamma \pots{}{\alpha\beta\gamma} \CF_{\beta\gamma}}{\sum_\beta\sum_\gamma \pots{}{\alpha\beta\gamma} \CF_{\beta\gamma}} - \frac{\sum_\gamma \pots{}{\alpha'\beta\gamma} \CF_{\beta\gamma}}{\sum_\beta\sum_\gamma \pots{}{\alpha'\beta\gamma} \CF_{\beta\gamma}}}.
  \end{equation*}
Interchanging the two suprema, fixing (for the moment) $\alpha$ and $\alpha'$,
defining $\tpots{}{\beta\gamma} := \pots{}{\alpha\beta\gamma} /
\pots{}{\alpha'\beta\gamma}$ and $\tCF_{\beta\gamma} := 
\CF_{\beta\gamma} \pots{}{\alpha'\beta\gamma}$, noting that we can without 
loss of generality assume that $\tCF$ is normalized in $\ell_1$ sense, the 
previous expression (apart from the $\frac{1}{2} \sup_{\alpha,\alpha'}$) 
simplifies to
  \begin{equation}\label{eq:sup_h}
  \sup_{\substack{\tCF>0,\\ \norms{\tCF}{1} = 1}} \sum_\beta \abs{\sum_\gamma \tCF_{\beta\gamma} \left(\frac{\tpots{}{\beta\gamma}}{\sum_\beta\sum_\gamma \tpots{}{\beta\gamma} \tCF_{\beta\gamma}} - 1\right)}.
  \end{equation}
In Appendix \ref{app:sup_h} we show that this equals
  \begin{equation}\label{eq:sup_h2}
  2 \sup_{\beta\ne\beta'} \sup_{\gamma,\gamma'} \tanh\left(\frac{1}{4}\log \frac{\tpots{}{\beta\gamma}}{\tpots{}{\beta'\gamma'}}\right).
  \end{equation}
We conclude that if we take all local norms to be the $\ell_\infty$ norms,
then $A_{\de{I}{i},\de{J}{j}}$ equals
\begin{equation}\begin{split}\label{eq:pot_strength_linfty}
  & N(\pot{I},i,j) \\
  & := 
\sup_{\alpha\ne\alpha'} \sup_{\beta\ne\beta'} \sup_{\gamma,\gamma'} \tanh \left( \frac{1}{4} \log \frac{\pots{I}{\alpha\beta\gamma}}{\pots{I}{\alpha'\beta\gamma}} \frac{\pots{I}{\alpha'\beta'\gamma'}}{\pots{I}{\alpha\beta'\gamma'}} \right),
\end{split}\end{equation}
which is defined for $i,j \in I$ with $i\ne j$ and where 
$\pots{I}{\alpha\beta\gamma}$ is shorthand for 
$\pot{I}(\rv{i} = \alpha,\rv{j}=\beta,\rv{I\setminus \{i,j\}}=\gamma)$; see 
Fig.\ \ref{fig:pot_strength} for an illustration. 

\begin{figure}[bt]
\centering
\psset{unit=0.7cm}
\psset{arrowscale=1.5}
\begin{pspicture}(-0.6,-2.6)(5,0.8)
{\small
\pscircle(0.0,0){0.3}
\rput(0.0,0.0){$i$}
\rput[B](0.0,0.5){$\rv{i} = \alpha$}
\rput(1.5,0.0){$I$}
\rput[B](1.5,0.5){$\pot{I}$}
\rput(3.0,0.0){$j$}
\rput[B](3.0,0.5){$\rv{j} = \beta$}
\rput(4.5,0.0){$J$}
\rput[B](1.5,-1.7){$\underbrace{\hskip 1.4cm}_{\displaystyle I\setminus \{i,j\}}$}
\rput[lB](2.7,-1.4){$\rv{I\setminus \{i,j\}} = \gamma$}
\psline(0.3,0.0)(1.2,0.0)
\psline{<-}(0.6,0.0)(1.2,0.0)
\psframe(1.2,0.3)(1.8,-0.3)
\psline(1.8,0.0)(2.7,0.0)
\pscircle(3.0,0.0){0.3}
\psframe(4.2,0.3)(4.8,-0.3)
\psline(3.3,0.0)(4.2,0.0)
\psline{<-}(3.6,0.0)(4.2,0.0)
\psline(1.4,-0.3)(0.8,-1.2)
\psline(1.5,-0.3)(1.5,-1.2)
\psline(1.6,-0.3)(2.2,-1.2)
\pscircle(0.8,-1.5){0.3}
\pscircle(1.5,-1.5){0.3}
\pscircle(2.2,-1.5){0.3}
}
\end{pspicture}
\caption{\label{fig:pot_strength}
Part of the factor graph relevant in expressions \eref{eq:pot_strength_linfty}, \eref{eq:sc_l1_linfty} and \eref{eq:sc_spectral_linfty}. Here $i,j \in I$ with $i\ne j$, and $J \in \nbe{j}{I}$.
}
\end{figure}  

Now combining \eref{eq:me_deriv_qmnorm}, \eref{eq:me_deriv_A} and
\eref{eq:pot_strength_linfty}, we finally obtain:
  \begin{equation*}
  \qnorm{f'\big(\quo{\lM}\big)} = \qnorm{f'(\lM)} \le 
  \max_{\de{J}{j}} \sum_{I\in\nbve{j}{J}} \sum_{i\in\nbfe{I}{j}} N(\pot{I},i,j).
  \end{equation*}
Applying Lemma \ref{lemm:sc_general} now yields that $\quo{f}$ is a 
contraction with respect to the quotient norm on $\lV / \lW$ if the
right-hand side is strictly smaller than 1.

Consider the mapping $\eta : \lV / \lW \to \lV$ that maps $\quo{\lM}$ to the
\emph{normalized} $\lM \in \lV$, i.e.\ such that 
$\norms{\exp \lme{I}{i}}{1} = 1$ for all components $\de{I}{i}$.
If we take for $f$ the $\ell_1$-normalized LBP update (in the log-domain),
the following diagram commutes:
  \begin{equation*}
  \begin{CD}
  \lV         @>f>>        \lV\\
  @VV{\pi}V                @AA{\eta}A \\
  \lV/\lW     @>\quo{f}>>  \lV/\lW
  \end{CD}
  \qquad\qquad f = \eta \circ \quo{f} \circ \pi.
  \end{equation*}
Since both $\pi$ and $\eta$ are continuous, we can translate convergence 
results for $\quo{f}$ back to similar results for $f$. We have proved:

\begin{theo}\label{theo:sc_l1_linfty}
If 
  \begin{equation}\label{eq:sc_l1_linfty}
  \max_{\de{J}{j}} \sum_{I\in\nbve{j}{J}} \sum_{i\in\nbfe{I}{j}} N(\pot{I},i,j) < 1,
\end{equation}
LBP converges to a unique fixed point irrespective of the initial messages.
\QED
\end{theo}

Now we can also generalize Corollary \ref{coro:sc_spectral_binary}:

\begin{theo}\label{theo:sc_spectral_linfty}
If the spectral radius of the matrix
  \begin{equation}\label{eq:sc_spectral_linfty}
  A_{\de{I}{i},\de{J}{j}} = \ind_{\nbve{j}{I}}(J) \ind_{\nbfe{I}{i}}(j) N(\pot{I},i,j),
\end{equation}
is strictly smaller than 1, LBP converges to a unique fixed point irrespective 
of the initial messages.
\end{theo}
\begin{proof}
Similar to the binary pairwise case; see Theorem \ref{theo:almost_Jacobi} in
Appendix \ref{app:quotients} for details.
\end{proof}

Note that Theorem \ref{theo:sc_l1_linfty} is a trivial consequence of Theorem
\ref{theo:sc_spectral_linfty}, since the $\ell_1$-norm is an upper bound on the
spectral radius. However, to prove the latter, it seems that we have to go
through all the work (and some more) needed to prove the former.

\subsection{Special cases}

In this subsection we study the implications for two special cases, namely
factor graphs that contain no cycles and the case of pairwise interactions.

\subsubsection{Trees}

Theorem \ref{theo:sc_spectral_linfty} gives us a proof of the well-known fact
that LBP converges on trees (whereas Theorem \ref{theo:sc_l1_linfty} is not
strong enough to prove that result):

\begin{coro}\label{coro:sc_tree}
If the factor graph is a tree, LBP converges to a
unique fixed point irrespective of the initial messages.
\end{coro}
\begin{proof}
The spectral radius of \eref{eq:sc_spectral_linfty} is easily shown to be zero
in this special case, for any choice of the potentials. 
\end{proof}

\subsubsection{Pairwise interactions}

We formulate Theorems \ref{theo:sc_l1_linfty} and \ref{theo:sc_spectral_linfty}
for the special case of pairwise interactions (which corresponds to $\gamma$ 
taking on only one value), i.e.\ all factors consists of either one or two 
variables. For a pair-potential $\pot{ij} = \pots{ij}{\alpha\beta}$, expression
\eref{eq:pot_strength_linfty} simplifies to (see also Fig.\ \ref{fig:pairwise})
\begin{equation}\label{eq:pot_strength_linfty_pairwise} N(\pot{ij}) :=
\sup_{\alpha\ne\alpha'} 
\sup_{\beta\ne\beta'}
\tanh \left( \frac{1}{4} \left( \log \frac{\pots{ij}{\alpha\beta}}{\pots{ij}{\alpha'\beta}} \frac{\pots{ij}{\alpha'\beta'}}{\pots{ij}{\alpha\beta'}} \right) \right).
\end{equation} 
Note that this quantity is invariant to ``reallocation'' of
single variable factors $\pot{i}$ or $\pot{j}$ to the pair factor $\pot{ij}$
(i.e.\ $N(\pot{ij}) = N(\pot{ij}\pot{i}\pot{j})$). $N(\pot{ij})$ can be
regarded as a measure of the strength of the potential $\pot{ij}$.

\begin{figure}[bt]
\centering
\psset{unit=0.7cm}
\psset{arrowscale=1.5}
\begin{pspicture}(-0.5,-1.8)(5.5,0.7)
{\small
\pscircle(0.0,0){0.3}
\rput(0.0,0.0){$k$}
\rput(0.0,-1.5){$k$}
\rput(0.7,-1.5){$\pot{k}$}
\rput(1.5,0.7){$\pot{ki}$}
\rput(1.5,0.0){$ki$}
\rput(3.0,0.0){$i$}
\rput(3.0,-1.5){$i$}
\rput(3.7,-1.5){$\pot{i}$}
\rput(4.5,0.7){$\pot{ij}$}
\rput(4.5,0.0){$ij$}
\rput(6.0,0.0){$j$}
\rput(6.0,-1.5){$j$}
\rput(6.7,-1.5){$\pot{j}$}
\psline(0.0,-0.3)(0.0,-1.2)
\psframe(-0.3,-1.2)(0.3,-1.8)
\psline(0.3,0.0)(1.2,0.0)
\psframe(1.2,-0.3)(1.8,0.3)
\psline(1.8,0.0)(2.7,0.0)
\psline{->}(1.8,0.0)(2.4,0.0)
\pscircle(3.0,0.0){0.3}
\psline(3.0,-0.3)(3.0,-1.2)
\psframe(2.7,-1.2)(3.3,-1.8)
\psline(3.3,0.0)(4.2,0.0)
\psframe(4.2,-0.3)(4.8,0.3)
\psline(4.8,0.0)(5.7,0.0)
\psline{->}(4.8,0.0)(5.4,0.0)
\pscircle(6.0,0.0){0.3}
\psline(6.0,-0.3)(6.0,-1.2)
\psframe(5.7,-1.2)(6.3,-1.8)
}
\end{pspicture}
\caption{Part of the factor graph in the pairwise case relevant in
\eref{eq:pot_strength_linfty_pairwise} and \eref{eq:sc_l1_linfty_pairwise}.
Here $k \in \del{i}$ and $j\in\dele{i}{k}$.
\label{fig:pairwise}
}
\end{figure}  

The $\ell_1$-norm based condition \eref{eq:sc_l1_linfty} can be written in the pairwise case as:
  \begin{equation}\label{eq:sc_l1_linfty_pairwise}
  \max_{i\in \ve} \max_{k\in \del{i}} \sum_{j\in \dele{i}{k}} N(\pot{ij}) < 1.
  \end{equation}
The matrix defined in \eref{eq:sc_spectral_linfty}, relevant for the spectral radius condition,
can be replaced by the following $\nel{\des}\times\nel{\des}$-matrix in the pairwise case:
  \begin{equation}\label{eq:sc_spectral_linfty_pairwise}
  A_{\de{i}{j},\de{k}{l}} := N(\pot{ij}) \delta_{i,l} \ind_{\dele{i}{j}}(k).
  \end{equation}

For the binary case, we reobtain our earlier results, since  
$N\big(\exp({J_{ij}\rv{i}\rv{j}})\big) = \tanh \abs{J_{ij}}$.

\subsection{Factors containing zeros}\label{sec:factors_zeros}

Until now, we have assumed that all factors are strictly positive. In many
interesting applications of the Sum-Product Algorithm, this assumption is
violated: the factors may contain zeros. It is thus interesting to see if and
how our results can be extended towards this more general case.

The easiest way to extend the results is by assuming that---although the
factors may contain zeros---the messages are guaranteed to remain strictly
positive (i.e.\ the log-messages remain finite) after each
update.\footnote{Additionally, the initial messages are required to be strictly
positive, but this requirement is easily met and is necessary for
obtaining good LBP results.} Even more general extensions with milder
conditions may exist, but we believe that considerably more work would be
required to overcome the technical problems that arise due to messages 
containing zeros. 

Assume that each factor $\pot{I}$ is a nonnegative function $\pot{I} :
\prod_{i\in I} \drv{i} \to [0,\infty)$. In addition, assume that all factors
involving only a single variable are strictly positive. This can be assumed
without loss of generality, since the single variable factors that contain one
or more zeros can simply be absorbed into multi-variable factors involving
the same variable. Additionally, for each $I \in \fas$ 
consisting of more than one variable, assume that
\begin{equation}\label{eq:condition_factors_zeros}
\forall_{i\in I} \, \forall_{\rv{i}\in\drv{i}} \exists_{\rv{\nbfe{I}{i}}\in\drv{\nbfe{I}{i}}} : \pot{I}(\rv{i},\rv{\nbfe{I}{i}}) > 0.
\end{equation}
These conditions guarantee that strictly positive messages remain strictly
positive under the update equations \eref{eq:me_update}, as one easily checks,
implying that we can still use the logarithmic parameterization of the
messages and that the derivative \eref{eq:me_deriv} is still well-defined.

The expression for the potential strength \eref{eq:pot_strength_linfty} can
be written in a way that is also well-defined if the potential $\pot{I}$ 
contains zeros:
\begin{equation}\begin{split}\label{eq:pot_strength_linfty_zeros}
  & N(\pot{I},i,j) \\
  & := \sup_{\alpha\ne\alpha'} \sup_{\beta\ne\beta'} \sup_{\gamma,\gamma'} 
\frac{\sqrt{\pots{I}{\alpha\beta\gamma} \pots{I}{\alpha'\beta'\gamma'}} - \sqrt{\pots{I}{\alpha'\beta\gamma} \pots{I}{\alpha\beta'\gamma'}}}{\sqrt{\pots{I}{\alpha\beta\gamma} \pots{I}{\alpha'\beta'\gamma'}} + \sqrt{\pots{I}{\alpha'\beta\gamma} \pots{I}{\alpha\beta'\gamma'}}}
\end{split}\end{equation}
which is defined for $i,j \in I$ with $i\ne j$ and where
$\pots{I}{\alpha\beta\gamma}$ is shorthand for
$\pot{I}(\rv{i} = \alpha,\rv{j}=\beta,\rv{I\setminus \{i,j\}}=\gamma)$.

The immediate generalization of Corollary \ref{theo:sc_spectral_linfty} is then as follows:
\begin{theo}\label{theo:sc_spectral_linfty_zeros}
Under the assumptions on the potentials described above (strict positivity
of single variable factors and \eref{eq:condition_factors_zeros} for
the other factors):
if the spectral radius of the matrix
  \begin{equation}\label{eq:sc_spectral_linfty_zeros}
  A_{\de{I}{i},\de{J}{j}} = \ind_{\nbve{j}{I}}(J) \ind_{\nbfe{I}{i}}(j) N(\pot{I},i,j),
\end{equation}
(with $N(\pot{I},i,j)$ defined in \eref{eq:pot_strength_linfty_zeros})
is strictly smaller than 1, LBP converges to a unique fixed point irrespective
of the initial messages.
\end{theo}
\begin{proof}
Similar to the strictly positive case. The only slight subtlety occurs in
Appendix \ref{app:sup_h} where one has to take a limit of strictly positive
factors converging to the desired nonnegative factor and use the continuity 
of the relevant expressions with respect to the factor entries to prove that the 
bound also holds in this limit.
\end{proof}

\subsubsection{Example}

Define, for $\epsilon \ge 0$, the (``ferromagnetic'') pair factor $\pot{}(\epsilon)$
by the following matrix:
\begin{equation*}
\pot{}(\epsilon) := \begin{pmatrix} 1 & \epsilon \\ \epsilon & 1\end{pmatrix}.
\end{equation*}
Now consider a binary pairwise factor graph, consisting of a single loop of $N$
binary variables, i.e.\ the network topology is that of a circle. Take for the
$N-1$ pair interactions $\pot{\{i,i+1\}}$ (for $i=1,2,\dots,N-1$) the identity
matrices (i.e.\ the above pair factors for $\epsilon = 0$) and take for the remaining 
one $\pot{\{1,N\}} = \pot{}(\epsilon)$ for some  $\epsilon \ge 0$.
Note that the potential strength 
$N(\pot{}(\epsilon)) = \frac{1-\epsilon}{1+\epsilon}$
converges to 1 as $\epsilon \downarrow 0$. The spectral radius of the corresponding
matrix $A_{\de{I}{i},\de{J}{j}}$ can be shown to be equal to 
$$\rho(A) = \left(\frac{1-\epsilon}{1+\epsilon}\right)^{1/N}$$
which is strictly smaller than 1 if and only if $\epsilon > 0$. Hence LBP converges
to a unique fixed point if $\epsilon > 0$. This result is sharp, since for 
$\epsilon = 0$, LBP simply ``rotates''
the messages around without changing them and hence no convergence occurs 
(except, obviously,
if the initial messages already correspond to the fixed point of uniform
messages).

\section{Comparison with other work}

In this section we explore the relations of our results with previously 
existing work.

\subsection{Comparison with work of Tatikonda and Jordan}

In \cite{TatikondaJordan02,Tatikonda03}, a connection is made between two
seemingly different topics, namely the Sum-Product Algorithm on the one hand
and the theory of Gibbs measures \cite{Georgii88} on the other hand. The main
result of \cite{TatikondaJordan02} states that LBP converges uniformly (to a
unique fixed point) if the Gibbs measure on the corresponding computation
tree\footnote{The computation tree is an ``unwrapping'' of the factor graph
with respect to the Sum-Product Algorithm; specifically, the computation tree
starting at variable $i \in \ve$ consists of all paths starting at $i$ that
never backtrack.} is unique.  

This is a remarkable and beautiful result; however, the question of convergence
of LBP is replaced by the question of uniqueness of the Gibbs measure, which
is not trivial. Fortunately, sufficient conditions for the uniqueness of the
Gibbs measure exist; the most well-known are \emph{Dobrushin's condition} and a
weaker (but easier verifiable) condition known as \emph{Simon's condition}. In
combination with the main result of \cite{TatikondaJordan02}, they yield
directly testable sufficient conditions for convergence of LBP to a unique
fixed point. For reference, we will state both results in our notation below.
For details, see \cite{TatikondaJordan02,Tatikonda03} and \cite{Georgii88}.
Note that the results are valid for the case of positive factors consisting of
at most two variables and that it is not obvious whether they can be
generalized.

\subsubsection{LBP convergence via Dobrushin's condition}

Define \emph{Dobrushin's interdependence matrix} as the $\nve \times \nve$ matrix $C$ with entries
\begin{equation}
C_{ij} := \sup_{\rv{\dele{i}{j}}} \sup_{\rv{j}, \rv{j}'} \frac{1}{2} \sum_{\rv{i}} \abs{ P(x_i \given x_{\dele{i}{j}}, x_j) - P(x_i \given x_{\dele{i}{j}}, x_j')} \\
\end{equation}
for $j \in \del{i}$ and 0 otherwise.
\begin{theo}\label{theo:Dobrushin}
For pairwise (positive) factors, LBP converges to a unique fixed point if
\begin{equation*}
\max_{i\in\ve} \sum_{j\in\del{i}} C_{ij} < 1.
\end{equation*}
\end{theo}
\begin{proof}
For a proof sketch, see \cite{Tatikonda03}. For the proof of Dobrushin's condition
see chapter 8 in \cite{Georgii88}.
\end{proof}

We can rewrite the conditional probabilities in terms of factors:
\begin{equation*}
P(x_i \given x_{\dele{i}{j}}, x_j)
= \frac{\potrv{i} \potrv{ij} \prod_{k\in\dele{i}{j}} \potrv{ik}}{\sum_{\rv{i}} \potrv{i} \potrv{ij} \prod_{k\in\dele{i}{j}} \potrv{ik}}.
\end{equation*}
Note that the complexity of the calculation of this quantity is generally 
exponential in the size of the neighborhood $\del{j}$, which may prohibit
practical application of Dobrushin's condition.

For the case of binary $\pm 1$-valued variables, some elementary algebraic manipulations yield
\begin{equation*}\begin{split}
C_{ij} & = \sup_{\rv{\dele{i}{j}}} \frac{\sinh 2\abs{\bJ{i}{j}}}{\cosh 2\bJ{i}{j} + \cosh 2(\bt{i} + \sum_{k\in\dele{i}{j}} \rv{k} \bJ{i}{k})} \\
& = \frac{\tanh (\abs{\bJ{i}{j}} - H_{ij}) + \tanh (\abs{\bJ{i}{j}} + H_{ij})}{2}
\end{split}\end{equation*}
with
\begin{equation*}
H_{ij} := \inf_{\rv{\dele{i}{j}}} \abs{\bt{i} + \sum_{k\in\dele{i}{j}} \rv{k} \bJ{i}{k}}.
\end{equation*}


\subsubsection{LBP convergence via Simon's condition}

Simon's condition is a sufficient condition for Dobrushin's condition (see proposition
8.8 in \cite{Georgii88}). This leads to a looser, but more easily verifiable, bound:

\begin{theo}\label{theo:Simon}
For pairwise (positive) factors, LBP converges to a unique fixed point if
\begin{equation*}
\max_{i\in\ve} \sum_{j\in\del{i}} \left( \frac{1}{2} \sup_{\alpha,\alpha'}\sup_{\beta,\beta'} \log \frac{\pots{ij}{\alpha\beta}}{\pots{ij}{\alpha'\beta'}} \right) < 1.
\end{equation*}
\hfill\QED
\end{theo}
It is not difficult to show that this bound is weaker than
\eref{eq:sc_l1_linfty_pairwise}. Furthermore,
unlike Dobrushin's condition and Corollary
\ref{coro:sc_spectral_improved_binary}, it does not take into account single
variable factors.

\subsection{Comparison with work of Ihler \emph{et al.}}\label{sec:ihler_discussion}

In the recent and independent work \cite{IhlerFisherWillsky04} of Ihler
\emph{et al.}, a methodology was used which is very similar to the one used in
this work. In particular, the same local $\ell_\infty$ quotient metric is used
to derive sufficient conditions for LBP to be a contraction. In the work
presented here, the Mean Value Theorem (in the form of Lemma
\ref{lemm:mean_value_coro}) is used in combination with a bound on the
derivative in order to obtain a bound on the convergence rate $K$ in
\eref{eq:contraction}. In contrast, in \cite{IhlerFisherWillsky04} a direct
bound on the distance of two outgoing messages is derived in terms of the
distance of two different products of incoming messages (equation (13) in
\cite{IhlerFisherWillsky04}). This bound becomes relatively stronger as the
distance of the products of incoming messages increases. This has the advantage
that it can lead to stronger conclusions about the effect of finite message 
perturbations than would be possible with our bound, based on the Mean Value 
Theorem. However, for the question of \emph{convergence}, the relevant limit 
turns out to be that of \emph{infinitesimal} message perturbations, i.e.\ it 
suffices to study the derivative of the LBP updates as we have done here. 

In the limit of infinitesimal message perturbations, the fundamental bound 
(13) in \cite{IhlerFisherWillsky04} leads to the following measure of potential strength:
  \begin{equation*}
  D(\pot{ij}) := \tanh \left( \frac{1}{2} \left( \sup_{\alpha,\beta} \sup_{\alpha',\beta'} \log \frac{\pots{ij}{\alpha\beta}}{\pots{ij}{\alpha'\beta'}} \right) \right).
  \end{equation*}
Using this measure, Ihler \emph{et.\ al} derive two different conditions for 
convergence of LBP. The first one is similar to our \eref{eq:sc_l1_linfty_pairwise}
and the second condition is equivalent to our spectral radius result 
\eref{eq:sc_spectral_linfty_pairwise}, except that in both conditions,
$N(\pot{ij})$ is used instead of $D(\pot{ij})$. 
The latter condition is formulated in \cite{IhlerFisherWillsky04} in terms 
of the convergence properties of an iterative BP-like algorithm. The 
equivalence of this formulation with a formulation in terms of the 
spectral radius of a matrix can be seen from the fact that for any square 
matrix $A$, $\rho(A) < 1$ if and only if $\lim_{n\to\infty} A^n = 0$. However,
our result also gives a contraction rate, unlike the iterative formulation in
\cite{IhlerFisherWillsky04}.

Thus, the results in \cite{IhlerFisherWillsky04} are similar to ours in the
pairwise case, except for the occurrence of $D(\pot{ij})$ instead of
$N(\pot{ij})$. It is not difficult to see that $N(\pot{ij}) \le D(\pot{ij})$
for any pair factor $\pot{ij}$; indeed, for any choice of
$\alpha,\beta,\gamma,\delta$:
  \begin{equation*}
  \sqrt{\pots{}{\alpha\gamma}\pots{}{\beta\delta}} \Big/ \sqrt{\pots{}{\beta\gamma}\pots{}{\alpha\delta}} \quad\le\quad \Big(\sup_{\sigma\tau}{\pots{}{\sigma\tau}}\Big) \Big/ \Big(\inf_{\sigma\tau}{\pots{}{\sigma\tau}} \Big).
  \end{equation*}
Thus the convergence results in \cite{IhlerFisherWillsky04} are similar to,
but weaker than the results derived in the present work.

After initial submission of this work, \cite{IhlerFisherWillsky05} was
published, which improves upon \cite{IhlerFisherWillsky04} by
exploiting the freedom of choice of the single node factors (which can be
``absorbed'' to an arbitrary amount by corresponding pair factors). This leads to an
improved measure of potential strength, which turns out to be identical to
our measure $N(\pot{ij})$. Thus, for pairwise, strictly positive potentials,
the results in \cite{IhlerFisherWillsky05} are equivalent to the results
\eref{eq:sc_l1_linfty_pairwise} and \eref{eq:sc_spectral_linfty_pairwise} 
presented here. Our most general results, Theorems
\ref{theo:sc_l1_linfty}, \ref{theo:sc_spectral_linfty} and
\ref{theo:sc_spectral_linfty_zeros} and Corollary
\ref{coro:sc_spectral_improved_binary}, are not present in
\cite{IhlerFisherWillsky05}.

\subsection{Comparison with work of Heskes}

A completely different methodology to obtain sufficient conditions for the
uniqueness of the LBP fixed point is used in \cite{Heskes04}. By studying the
Bethe free energy and exploiting the relationship between properties of the
Bethe free energy and the LBP algorithm, conclusions are drawn about the
uniqueness of the LBP fixed point; however, whether uniqueness of the fixed
point also implies convergence of LBP seems to be an open question. We 
state the main result of \cite{Heskes04} in our notation below. 

The following measure of potential strength is used in \cite{Heskes04}. For $I \in \fas$, let
\begin{equation*}\begin{split}
\omega_{I} := \sup_{\rv{I}} \sup_{\rv{I}'} \Big( & \log \potrv{I} + (\nel{I} - 1)\log \pot{I}(\rv{I}') \\
& {} - \sum_{i\in I} \log \pot{I}(\rv{I\setminus i}',\rv{i}) \Big).
\end{split}\end{equation*}
The potential strength is then defined as $\sigma_{I} := 1 - e^{-\omega_I}$. 

\begin{theo}\label{theo:Heskes}
LBP has a unique fixed point if there exists an ``allocation matrix'' $X_{Ii}$ between
factors $I \in \fas$ and variables $i \in \ve$ such that
\begin{enumerate}
\item $X_{Ii} \ge 0$ for all $I \in \fas, i \in I$;
\item $(1 - \sigma_I) \max_{i \in I} X_{Ii} + \sigma_I \sum_{i \in I} X_{Ii} \le 1$ for all $I \in \fas$;
\item $\sum_{I\in N_i} X_{Ii} \ge \nel{N_i} - 1$ for all $i \in \ve$.
\end{enumerate}
\end{theo}
\begin{proof}
See Theorem 8.1 in \cite{Heskes04}.
\end{proof}
The (non)existence of such a matrix can be determined using standard linear 
programming techniques.


\section{Numerical comparison of various bounds}\label{sec:bound_comparison}

In this subsection, we compare various bounds on binary pairwise graphical
models, defined in \eref{eq:probability_measure_binary}, for various choices of
the parameters.  First we study the case of a completely uniform model (i.e.\
full connectivity, uniform couplings and uniform local fields). Then we study
non-uniform couplings $J_{ij}$, in the absence of local fields. Finally, we 
take fully random models in various parameter regimes (weak/strong local fields,
strong/weak ferromagnetic/spin-glass/anti-ferromagnetic couplings).

\subsection{Uniform couplings, uniform local field}\label{sec:compare}

The fully connected Ising model consisting of $N$ binary $\pm1$-valued
variables with uniform couplings $J$ and uniform local field $\theta$ is
special in the sense that an exact description of the parameter region for
which the Gibbs measure on the computation tree is unique, is available. Using
the results of Tatikonda and Jordan, this yields a strong bound on the
parameter region for which LBP converges to a unique fixed point.  Indeed, the
corresponding computation tree is a uniform Ising model on a Cayley tree of
degree $N-2$, for which \mbox{(semi-)}analytical expressions for the
paramagnetic--ferromagnetic and paramagnetic--anti-ferromagnetic phase
transition boundaries are known (see section 12.2 in \cite{Georgii88}). Since
the Gibbs measure is known to be unique in the paramagnetic phase, this gives an exact
description of the $(J,\theta)$ region for which the Gibbs measure on the
computation tree is unique, and hence a bound on LBP convergence on the original model.

\begin{figure}[bt]
\centering
\includegraphics[width=\columnwidth]{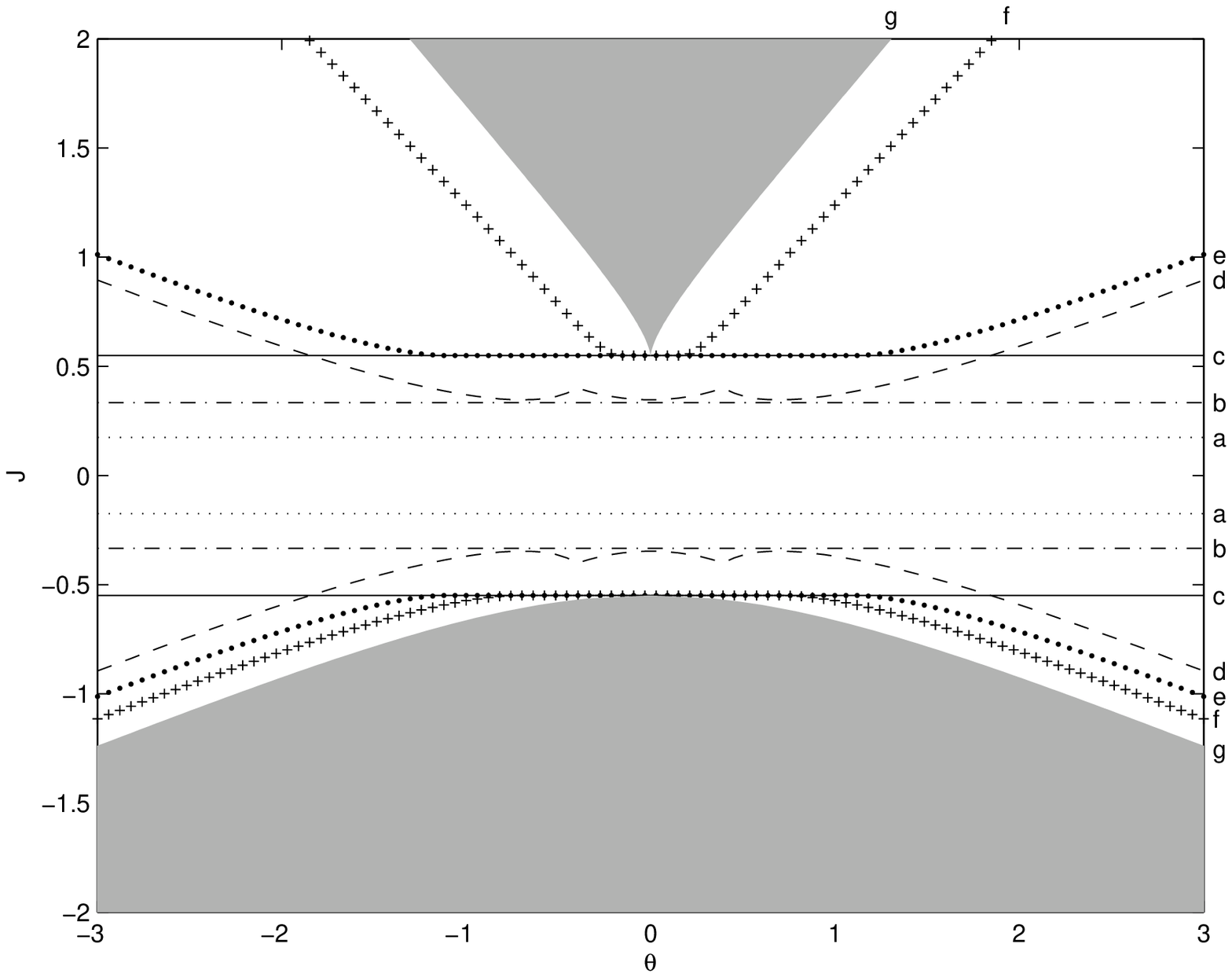}
\caption{\label{fig:compare}
Comparison of various LBP convergence bounds for the fully connected $N = 4$ binary
Ising model with uniform coupling $J$ and uniform local field $\theta$. 
(a) Heskes' condition (b) Simon's condition (c) spectral radius condition
(d) Dobrushin's condition (e) improved spectral radius condition for $m=1$
(f) improved spectral radius condition for $m=5$ (g) 
uniqueness of Gibbs' measure condition. See the main text (section 
\ref{sec:compare}) for more explanation.}
\end{figure}  

In Fig.~\ref{fig:compare} we have plotted various bounds on LBP
convergence in the $(J,\theta)$ plane for $N = 4$ (other values of $N$ yield
qualitatively similar results). The gray area (g) marks regions where the
Gibbs measure on the computation tree is not unique; in the white area, the
Gibbs measure is unique and hence LBP is guaranteed to converge. Note that this
bound is only available due to the high symmetry of the model.
In \cite{TagaMase06} it is shown that parallel LBP does not converge in the
lower (anti-ferromagnetic) gray region. In the upper (ferromagnetic) region
on the other hand, parallel LBP does converge, but it may be that the fixed
point is no longer unique.

The various lines correspond to different sufficient conditions for LBP
convergence; the regions enclosed by two lines of the same type (i.e.\ the
inner regions for which $J$ is small) mark the regions of guaranteed
convergence. The lightly dotted lines (a) correspond with Heskes' condition,
Theorem \ref{theo:Heskes}. The dash-dotted lines (b)
correspond with Simon's condition, Theorem \ref{theo:Simon}. The dashed lines
(d) correspond with Dobrushin's condition (Theorem \ref{theo:Dobrushin}), 
which is seen to improve upon Simon's condition for $\theta \ne 0$, but is 
nowhere sharp. The solid lines (c) correspond with the spectral radius condition
Corollary \ref{coro:sc_spectral_binary} (which coincides with the $\ell_1$-norm 
based condition Corollary \ref{coro:sc_l1_binary} in this case and is also
equivalent to the result of \cite{IhlerFisherWillsky04}), which is independent 
of $\theta$ but is actually sharp for $\theta = 0$. The heavily dotted
lines (e) correspond to Corollary \ref{coro:sc_spectral_improved_binary} with
$m = 1$, the $+$-shaped lines (f) to the same Corollary with $m = 5$. Both
(e) and (f) are seen to coincide with (c) for small $\theta$, but improve for
large $\theta$.

We conclude that the presence of local fields makes it more difficult to obtain
sharp bounds on LBP convergence; only Dobrushin's condition (Theorem
\ref{theo:Dobrushin}) and Corollary \ref{coro:sc_spectral_improved_binary} take
into account local fields. Furthermore, in this case, our result Corollary
\ref{coro:sc_spectral_improved_binary} is stronger than the other bounds. Note
that the calculation of Dobrushin's condition is exponential in the number of
variables $\nve$, whereas the time complexity of our bound is polynomial in $\nve$.
Similar results are obtained for higher values of $N$.

\subsection{Non-uniform couplings, zero local fields}\label{sec:compare2}

\begin{figure}[bt]
\centering
\includegraphics[width=\columnwidth]{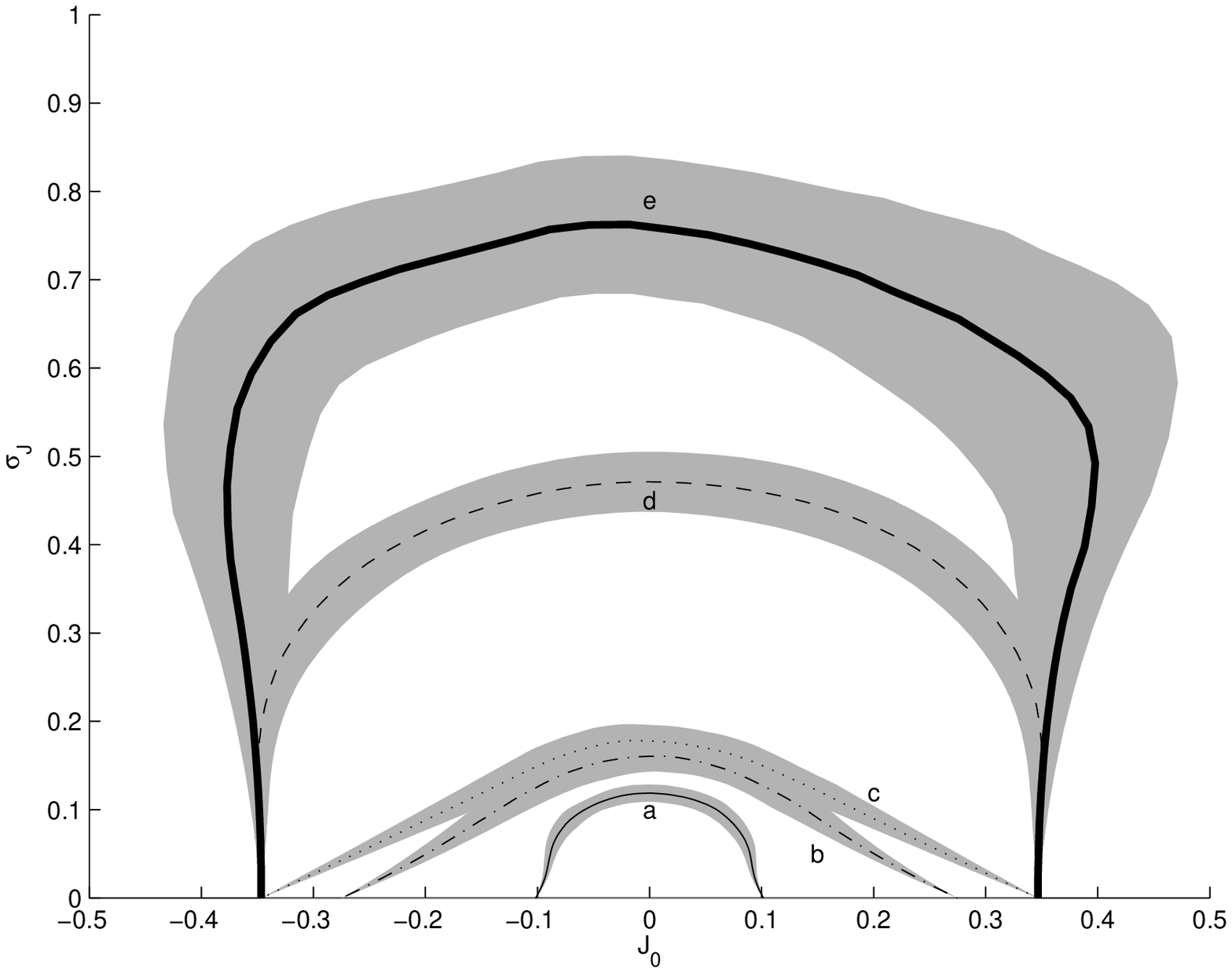}
\caption{\label{fig:compare2}
Comparison of various bounds for LBP convergence for toroidal Ising model
of size $10\times10$ with normally distributed couplings $\bJ{i}{j} \sim \C{N}(J_0, \sigma_J)$
and zero local fields. (a) Heskes' condition (b) Dobrushin's condition 
(c) $\ell_1$-norm condition (d) spectral radius condition (e) empirical 
convergence. See the main text (section \ref{sec:compare2}) for more 
explanation.}
\end{figure}

We have investigated in more detail the influence of the distribution of the
couplings $\bJ{i}{j}$, in the absence of local fields, and have also compared
with the empirical convergence behavior of LBP. We have taken a binary Ising
model on a rectangular toroidal grid (i.e.\ with periodic boundary conditions)
of size $10\times10$. The couplings were random independent normally
distributed nearest-neighbor couplings $\bJ{i}{j} \sim \C{N}(J_0, \sigma_J)$,
the local fields were $\bt{i} = 0$. Let $(r_J, \phi_J)$ be the polar
coordinates corresponding to the Cartesian coordinates $(J_0, \sigma_J)$. For
various angles $\phi_J \in [0,\pi]$, we have determined the critical radius
$r_J$ for each bound. The results have been averaged over 40 instances of the
model and can be found in Fig.~\ref{fig:compare2}. The lines correspond to
the mean bounds, the gray areas are ``error bars'' of one standard deviation.
The inner area (for which the couplings are small) bounded by each line means
``convergence'', either guaranteed or empirical (thus the larger the enclosed
area, the tighter the bound). From bottom to top: the thin solid line (a)
corresponds with Heskes' result (Theorem \ref{theo:Heskes}), the dash-dotted line
(b) with Dobrushin's condition (Theorem \ref{theo:Dobrushin}), the dotted line
(c) corresponds with the $\ell_1$-norm based condition Corollary
\ref{coro:sc_l1_binary}, the dashed line (d) with the spectral radius condition
Corollary \ref{coro:sc_spectral_binary} and the thick solid line (e) with the
empirical convergence behavior of LBP. 


We conclude from Fig.~\ref{fig:compare2} that the spectral radius condition
improves upon the $\ell_1$-norm based condition for non-uniform couplings and that
the improvement can be quite substantial. For uniform couplings (and zero local
fields), both conditions coincide and it can be proved that they are sharp 
\cite{MooijKappen05}.

\subsection{Fully random models}

Finally, we have considered fully connected binary pairwise graphical models
with completely random couplings and local fields (in various parameter regimes). 
We drew random couplings and local fields as follows: first, we drew i.i.d.\ 
random parameters $J_0, \sigma_J, \theta_0, \sigma_\theta$ from a normal 
distribution with mean 0 and variance 1. Then, for each variable $i$ we 
independently drew a local field parameter $\bt{i} \sim \C{N}(\theta_0,\sigma_\theta)$, 
and for each pair $\{i,j\}$ we independently drew a coupling parameter $\bJ{i}{j} \sim \C{N}(J_0,\sigma_J)$.

For the resulting graphical model, we have verified whether various sufficient
conditions for LBP convergence hold. If condition A holds whereas condition B
does not hold, we say that A wins from B. We have counted for each ordered pair
$(A,B)$ of conditions how often A wins from B. The results (for 50000 random
models consisting of $N = 4,8$ variables) can be found in Table
\ref{tab:comparison}: the number at row $A$, column $B$ is the number of trials
for which bound $A$ wins from bound $B$. On the diagonal ($A = B$) is the total
number of trials for which bound $A$ predicts convergence.  Theorem
\ref{theo:Dobrushin} is due to \cite{Tatikonda03}, Corollary 
\ref{coro:sc_spectral_binary} was first published (for the binary case) in 
\cite{IhlerFisherWillsky04} and Theorem \ref{theo:Heskes} is due to \cite{Heskes04}.

\begin{table}
\renewcommand{\arraystretch}{1.3}
\caption{Comparison of bounds (50000 trials, for $N = 4$ and $N = 8$)}
\label{tab:comparison}
\centering
\begin{tabular}{l|rrrr}
\bfseries $N = 4$                            & Th.\ \ref{theo:Dobrushin} & Cor.\ \ref{coro:sc_spectral_binary} & Th.\ \ref{theo:Heskes} & Cor.\ \ref{coro:sc_spectral_improved_binary} \\
\hline
Th.\ \ref{theo:Dobrushin}, \cite{Tatikonda03}                     &      (5779) &        170 &       3564 &          0 \\ 
Cor.\ \ref{coro:sc_spectral_binary}, \cite{IhlerFisherWillsky04}  &       10849 &    (16458) &      13905 &          0 \\ 
Th.\ \ref{theo:Heskes}, \cite{Heskes04}                           &         338 &          0 &     (2553) &          0 \\ 
Cor.\ \ref{coro:sc_spectral_improved_binary}, $m=1$, this work           &       13820 &       3141 &      17046 &    (19599) \\ 
\multicolumn{5}{c}{}\\
\multicolumn{5}{c}{}\\
\bfseries $N = 8$                           & Th.\ \ref{theo:Dobrushin} & Cor.\ \ref{coro:sc_spectral_binary} & Th.\ \ref{theo:Heskes} & Cor.\ \ref{coro:sc_spectral_improved_binary} \\
\hline
Th.\ \ref{theo:Dobrushin}, \cite{Tatikonda03}                     &       (668) &         39 &        597 &          0 \\ 
Cor.\ \ref{coro:sc_spectral_binary}, \cite{IhlerFisherWillsky04}  &         507 &     (1136) &       1065 &          0 \\ 
Th.\ \ref{theo:Heskes}, \cite{Heskes04}                           &           0 &          0 &       (71) &          0 \\ 
Cor.\ \ref{coro:sc_spectral_improved_binary}, $m=1$, this work           &         972 &        504 &       1569 &     (1640)    
\end{tabular}
\end{table}

Our result Corollary \ref{coro:sc_spectral_improved_binary} (for $m=1$)
outperforms the other bounds in each trial. For other values of $N$, we
obtain similar results.

\section{Discussion}\label{sec:discussion}

In this paper we have derived sufficient conditions for convergence of LBP to a
unique fixed point. Our conditions are directly applicable to arbitrary
graphical models with discrete variables and nonnegative factors. This is in
contrast with the sufficient conditions of Tatikonda and Jordan and with the
results of Ihler, Fisher and Willsky, which were only formulated for pairwise,
positive factors. We have shown cases where our results are stronger than
previously known sufficient conditions.

Our numerical experiments lead us to conjecture that Corollary 
\ref{coro:sc_spectral_improved_binary} is stronger
than the other bounds. We have no proof for this conjecture at the moment,
apart from the obvious fact that Corollary \ref{coro:sc_spectral_binary} is
weaker than Corollary \ref{coro:sc_spectral_improved_binary}. To prove that
Corollary \ref{coro:sc_spectral_improved_binary} is stronger than Theorem
\ref{theo:Dobrushin} seems subtle, since it is generally not the case that
$\rho(A) \le \norms{C}{\infty}$, although it seems that the weaker relation
$\norms{C}{\infty} < 1 \implies \rho(A) < 1$ does hold in general. The relation
with the condition in Theorem \ref{theo:Heskes} is not evident as well.

In the binary pairwise case, it turned out to be possible to derive sufficient
conditions that take into account local evidence (Corollary
\ref{coro:sc_spectral_improved_binary}). In the general case, such an
improvement is possible in principle but seems to be more involved. The
resulting optimization problem (essentially \eref{eq:sup_h} with additional
assumptions on $h$) looks difficult in general. If the variables' cardinalities
and connectivies are small, the resulting optimization problem can be solved,
but writing down a general solution does not appear to be trivial. The
question of finding an efficient solution in the general case is left for
future investigation. 

The work reported here raises new questions, some of which have been
(partially) answered elsewhere after the initial submission of this paper. The
influence of damping the LBP update equations has been considered for the
binary pairwise case in \cite{MooijKappen05c}, where it was shown that damping
has the most effect for anti-ferromagnetic interactions. Furthermore, it has
been proved in \cite{MooijKappen05c} that the bounds for LBP convergence
derived in the present work are sharp in the case of binary variables with
\mbox{(anti-)}ferromagnetic pairwise interactions and zero local fields, as
suggested by Fig.~\ref{fig:compare2}. An
extension of the results towards sequential update schemes has been given in
\cite{ElidanMcGrawKoller06}; it is shown that for each reasonable sequential
update scheme, the same conditions for convergence to a unique fixed point as
derived in this work apply. Likewise, in \cite{TagaMase06} it is shown that 
Dobrushin's condition is also valid for sequential LBP.

\useRomanappendicesfalse

\appendices

\renewcommand{\theequation}{\thesection.\arabic{equation}}

\section{Generalizing the $\ell_1$-norm\label{app:quotients}}

Let $(V_i,\norms{\cdot}{i})$ be a collection of normed vector spaces and
let $V = \bigoplus_i V_i$ be the direct sum of the $V_i$'s. 
The function $\norm{\cdot} : V \to \RN$ defined by
  \begin{equation}\label{eq:gen_l1_norm}
  \norm{v} := \sum_i \norms{v_i}{i}
  \end{equation}
is a norm on $V$, as one easily checks. Let $A : V \to
V$ be a linear mapping with ``blocks'' $A_{ij} : V_j \to V_i$ defined by
  \begin{equation*}
  \forall v_j \in V_j : \quad  Av_j = \sum_i A_{ij}v_j, \quad A_{ij}v_j \in V_i
  \end{equation*}
for all $j$.

\begin{theo}
The matrix norm of $A$ induced by the vector norm $\norm{\cdot}$ is given by:
  \begin{equation}\label{eq:gen_l1_mnorm}
  \norm{A} = \max_j \sum_i \mnorm{A_{ij}}{i}{j}
  \end{equation}
where
  \begin{equation*}
  \mnorm{A_{ij}}{i}{j} := \sup_{\substack{x \in V_j,\\ \norms{x}{j} \le 1}} \norms{A_{ij} x}{i}.
  \end{equation*}
\end{theo}
\begin{proof}
Let $v_k \in V_k$ such that $\norms{v_k}{k} = 1$. Then 
  \begin{equation*}\begin{split}
  \norm{Av_k} & = \norm{\sum_i A_{ik} v_k} = \sum_i \norms{A_{ik}v_k}{i} \\
  & \le \sum_i \mnorm{A_{ik}}{i}{k} \le \max_j \sum_i \mnorm{A_{ij}}{i}{j}.
  \end{split}\end{equation*}
Now let $v \in V$ such that $\norm{v}=1$. Then $v$ can be written as the convex
combination $v = \sum_k \norms{v_k}{k}\tilde v_k$, where
  \begin{equation*}
  \tilde v_k := \begin{cases}
     \frac{v_k}{\norms{v_k}{k}} & \text{if $v_k \ne 0$} \\
     0  			  & \text{if $v_k = 0$}.
  \end{cases}
  \end{equation*}
Hence:
  \begin{equation*}\begin{split}
  \norm{Av} & = \norm{\sum_k \norms{v_k}{k} A \tilde v_k} 
              \le \sum_k \norms{v_k}{k} \norm{A\tilde v_k} \\
            & \le \max_j \sum_i \mnorm{A_{ij}}{i}{j}. 
  \end{split}\end{equation*}
It is evident that this value is also achieved for some $v \in V$ with $\norm{v} = 1$.
\end{proof}

An illustrative example is obtained by considering $V=\RN^N$ to be the direct
sum of $N$ copies of $\RN$ with the absolute value as norm; then the norm
\eref{eq:gen_l1_norm} on $\RN^N$ is simply the $\ell_1$-norm and the induced
matrix norm \eref{eq:gen_l1_mnorm} reduces to \eref{eq:l1_mnorm}.

Suppose that each $V_i$ has a linear subspace $W_i$. We can consider the
quotient spaces $V_i / W_i$ with quotient norms $\qnorms{\cdot}{i}$.
The direct sum $W := \bigoplus_i W_i$ is itself a subspace of $V$, yielding a
quotient space $V/W$. For $v \in V$ we have $\quo{v} = \sum_i
\quo{v_i}$ and hence $V/W = \bigoplus_i (V_i / W_i)$. The quotient norm on
$V/W$ is simply the sum of the quotient norms on the $V_i / W_i$:
  \begin{equation}\label{eq:gen_l1_qnorm}\begin{split}
  \qnorm{v} 
  & := \inf_{w \in W} \norm{v+w} 
    = \inf_{w \in W} \sum_i \norms{v_i+w_i}{i} \\
  & = \sum_i \inf_{w_i \in W_i} \norms{v_i+w_i}{i}
    = \sum_i \qnorms{v_i}{i}.
  \end{split}\end{equation}

Let $A : V \to V$ be a linear mapping such that $AW \subseteq W$. Then $A$
induces a linear $\quo{A} : V/W \to V/W$; since $A_{ij} W_j \subseteq W_i$,
each block $A_{ij}: V_j \to V_i$ induces a linear $\quo{A_{ij}} : V_j/W_j \to
V_i/W_i$, and $\quo{A}$ can be regarded as consisting of the blocks 
$\quo{A_{ij}}$.

\begin{coro}
The matrix norm of $\quo{A} : V/W \to V/W$ induced by the quotient norm $\qnorm{\cdot}$ on $V/W$ is:
\begin{equation}\label{eq:gen_l1_qmnorm}
  \qnorm{A} 
   = \max_j \sum_i \mqnorm{A_{ij}}{i}{j}
  \end{equation}
where
  \begin{equation}\label{eq:gen_l1_lmqnorm}
  \mqnorm{A_{ij}}{i}{j} 
   = \sup_{\substack{x\in V_j,\\ \norms{x}{j}\le 1}} \qnorms{A_{ij}x}{i}.
  \end{equation}
\end{coro}
\begin{proof}
We can directly apply the previous Theorem to the quotient spaces to
obtain \eref{eq:gen_l1_qmnorm}; because 
  \begin{equation*}
  \{\quo{x} \in V_j/W_j: \qnorms{x}{j} \le 1\} = \quo{\{x\in V_j : \norms{x}{j} \le 1\}},
  \end{equation*}
we have:
  \begin{equation*}
  \mqnorm{A_{ij}}{i}{j} 
  := \sup_{\substack{\quo{x} \in V_j/W_j \\ \qnorms{x}{j} \le 1}} \norms{\quo{A_{ij}} \quo{x}}{i} 
  = \sup_{\substack{x \in V_j \\ \norms{x}{j} \le 1}} \qnorms{A_{ij}x}{i}.
  \end{equation*}
\end{proof}

For a linear $A : V \to V$ such that $AW\subseteq W$, we define the matrix
$\abs{A}_{ij}$ with entries $\abs{A}_{ij} := \mqnorm{A_{ij}}{i}{j}$.
Let $A,B$ be two such linear mappings; then
  \begin{equation*}\begin{split}
  \abs{AB}_{ij} 
  & = \mqnorm{(AB)_{ij}}{i}{j} 
    = \mnorm{\sum_k \quo{A_{ik} B_{kj}}}{i}{j} \\
  & \le \sum_k \mqnorm{A_{ik} \quo{B_{kj}}}{i}{j}
    \le \sum_k \mqnorm{A_{ik}}{i}{k} \mqnorm{B_{kj}}{k}{j} \\
  & = \sum_k \abs{A}_{ik} \abs{B}_{kj}
  \end{split}\end{equation*}
hence $\abs{AB} \le \abs{A}\abs{B}$. Note that 
$\norms{\abs{A}}{1} = \qnorm{A}$. We can generalize Theorem
\ref{theo:almost_Jacobi_binary}:

\begin{theo}\label{theo:almost_Jacobi}
Let $f : V \to V$ be differentiable and suppose that it satisfies 
\eref{eq:f_symmetry}. Suppose further that $\abs{f'(v)} \le A$ for some matrix
$A_{ij}$ (which does not depend on $v$) with $\rho(A) < 1$. Then for any 
$\quo{v}\in V/W$, the sequence 
$\quo{v}, \quo{f}(\quo{v}), \quo{f}^2(\quo{v}),\dots$ obtained by iterating 
$\quo{f}$ converges to a unique fixed point $\quo{v}_\infty$. 
\end{theo}
\begin{proof}
Using the chain rule, we have for any $n=1,2,\dots$ and any $v \in V$:
  \begin{equation*}\begin{split}
  & \norm{(\quo{f}^n)'(\quo{v})} 
    = \qnorm{(f^n)'(v)} 
    = \qnorm{\prod_{i=1}^{n} f'\big(f^{i-1}(v)\big)} \\
  &\quad = \norms{\abs{\prod_{i=1}^{n} f'\big(f^{i-1}(v)\big)}}{1} 
    \le \norms{\prod_{i=1}^{n} \abs{f'\big(f^{i-1}(v)\big)}}{1} \\
  &\quad  \le \norms{\prod_{i=1}^{n} A}{1} 
  = \norms{A^n}{1}.
  \end{split}\end{equation*}
By the Gelfand Spectral Radius Theorem, $\norms{A^n}{1}^{1/n} \to \rho(A)$
for $n\to\infty$. Choose $\epsilon>0$ such that $\rho(A) + \epsilon < 1$.
For some $N$, $\norms{A^N}{1} \le (\rho(A) + \epsilon)^N < 1$. Hence 
$\norm{(\quo{f}^N)'(\quo{v})} < 1$ for all $\quo{v} \in V/W$.
By Lemma \ref{lemm:sc_general}, $\quo{f}^N$ is a contraction with respect
to the quotient norm on $V / W$. Now apply Lemma \ref{lemm:pow_contract_conv}.
\end{proof}

\section{Proof that \eref{eq:sup_h} equals \eref{eq:sup_h2}\label{app:sup_h}}

Let $\tpots{}{\beta\gamma}$ be a matrix of positive numbers. Let 
  \begin{equation*}
  \C{H} := \{ \CF: \CF_{\beta\gamma} \ge 0, \sum_{\beta,\gamma} \CF_{\beta\gamma} = 1\}.
  \end{equation*}
Define the function $g : \C{H} \to \RN$ by
  \begin{equation*}
  g(\CF) = \sum_\beta \abs{\sum_\gamma \CF_{\beta\gamma} \left(\frac{\tpots{}{\beta\gamma}}{\sum_\beta\sum_\gamma \tpots{}{\beta\gamma} \CF_{\beta\gamma}} - 1\right)}.
  \end{equation*}

\begin{theo}
  \begin{equation*}
  \sup_{\CF \in \C{H}} g(h) = 2 \sup_{\beta\ne\beta'} \sup_{\gamma,\gamma'} \tanh\left(\frac{1}{4}\log \frac{\tpots{}{\beta\gamma}}{\tpots{}{\beta'\gamma'}}\right).
  \end{equation*}
\end{theo}
\begin{proof}
First note that we can assume without loss of generality that all 
$\tpots{}{\beta\gamma}$ are different, because of continuity. 
Define 
  \begin{align*}
  & \tpots{}{-} := \inf_{\beta\gamma} \tpots{}{\beta\gamma}, \qquad
  \tpots{}{+} := \sup_{\beta\gamma} \tpots{}{\beta\gamma}, \\
  & X := [\tpots{}{-},\tpots{}{+}], \qquad 
  X' := X \setminus \{\tpots{}{\beta\gamma} : \beta,\gamma\}.
  \end{align*}
For $\Psi \in X$, define
  \begin{equation*}
  \C{H}_\Psi := \{\CF \in \C{H} : \sum_{\beta,\gamma} \tpots{}{\beta\gamma} \CF_{\beta\gamma} = \Psi\},
  \end{equation*}
which is evidently a closed convex set. The function
  \begin{equation*}
  g_\Psi : \C{H}_\Psi \to \RN : \CF \mapsto \sum_\beta \abs{\sum_\gamma \CF_{\beta\gamma} \left(\frac{\tpots{}{\beta\gamma}}{\Psi} - 1\right)}
  \end{equation*}
obtained by restricting $g$ to $\C{H}_\Psi$ is convex. Hence it achieves 
its maximum on an extremal point of its domain.

Define 
  \begin{equation*}
  \C{H}_2 := \big\{h \in \C{H} : \#\{(\beta,\gamma) : h_{\beta\gamma} > 0\} = 2\big\}
  \end{equation*}
as those $h\in\C{H}$ with exactly two non-zero components. For $h \in \C{H}_2$,
define $\tpots{}{-}(h) := \inf\{\tpots{}{\beta\gamma} : h_{\beta\gamma} \ne
0\}$ and $\tpots{}{+}(h) := \sup\{\tpots{}{\beta\gamma} : h_{\beta\gamma} \ne
0\}$.  Because of continuity, we can restrict ourselves to the $\Psi \in X'$,
in which case the extremal points of $\C{H}_\Psi$ are precisely
$\C{H}_\Psi^* = \C{H}_\Psi \cap \C{H}_2$ (i.e.\ the extremal points have
exactly two non-zero components).

Now 
  \begin{equation*}\begin{split}
  \sup_{h \in \C{H}} g(h) 
  & = \sup_{\Psi\in X} \sup_{h \in \C{H}_\Psi} g_\Psi(h) \\
  & = \sup_{\Psi \in X'} \sup_{h \in \C{H}_\Psi^*} g_\Psi(h) \\
  & = \sup_{h \in \C{H}_2} \ \sup_{\tpots{}{-}(h) \le \Psi \le \tpots{}{+}(h)}\ g_\Psi(h) \\
  & = \sup_{h \in \C{H}_2} g(h).
  \end{split}\end{equation*}

For those $h \in \C{H}_2$ with components with different $\beta$, we can use
the Lemma below.
The $h \in \C{H}_2$ with components with equal $\beta$ are suboptimal, since
the two contributions in the sum over $\gamma$ in $g(h)$ have opposite sign.
Hence
  \begin{equation*}
  \sup_{\CF \in \C{H}_2} g(\CF) 
  = 2 \sup_{\beta\ne\beta'} \sup_{\gamma,\gamma'} \tanh\left(\frac{1}{4}\log \frac{\tpots{}{\beta\gamma}}{\tpots{}{\beta'\gamma'}}\right).
  \end{equation*}
\end{proof}

\begin{lemm}
Let $0 < a < b$. Then
  \begin{equation*}\begin{split}
  & \sup_{\substack{\eta \in (0,1)^2 \\ \eta_1 + \eta_2 = 1}} \eta_1 \abs{\frac{a}{\eta_1 a + \eta_2 b} - 1} + \eta_2 \abs{\frac{b}{\eta_1 a + \eta_2 b} - 1} \\
   & = 2 \tanh \left( \frac{1}{4} \log \frac{b}{a} \right) = 2 \frac{\sqrt{b} - \sqrt{a}}{\sqrt{b} + \sqrt{a}}.
  \end{split}\end{equation*}
\end{lemm}
\begin{proof}
Elementary. The easiest way to see this is to reparameterize
$\eta = (\frac{e^\nu}{2 \cosh\nu}, \frac{e^{-\nu}}{2 \cosh\nu})$ with $\nu \in (-\infty,\infty)$.
\end{proof}

\section*{Acknowledgment}
The research reported here is part of the Interactive Collaborative
Information Systems (ICIS) project (supported by the Dutch Ministry of Economic
Affairs, grant BSIK03024) and was also sponsored in part by the Dutch
Technology Foundation (STW). We thank Martijn Leisink for stimulating discussions.


\bibliographystyle{IEEEtran}


\begin{thebibliography}{10}
\providecommand{\url}[1]{#1}
\csname url@rmstyle\endcsname
\providecommand{\newblock}{\relax}
\providecommand{\bibinfo}[2]{#2}
\providecommand\BIBentrySTDinterwordspacing{\spaceskip=0pt\relax}
\providecommand\BIBentryALTinterwordstretchfactor{4}
\providecommand\BIBentryALTinterwordspacing{\spaceskip=\fontdimen2\font plus
\BIBentryALTinterwordstretchfactor\fontdimen3\font minus
  \fontdimen4\font\relax}
\providecommand\BIBforeignlanguage[2]{{%
\expandafter\ifx\csname l@#1\endcsname\relax
\typeout{** WARNING: IEEEtran.bst: No hyphenation pattern has been}%
\typeout{** loaded for the language `#1'. Using the pattern for}%
\typeout{** the default language instead.}%
\else
\language=\csname l@#1\endcsname
\fi
#2}}

\bibitem{MooijKappen05}
J.~M. Mooij and H.~J. Kappen, ``Sufficient conditions for convergence of loopy
  belief propagation,'' in \emph{Proc. of the 21st Annual Conf. on Uncertainty
  in Artificial Intelligence (UAI-05)}.\hskip 1em plus 0.5em minus 0.4em\relax
  Corvallis, Oregon: AUAI Press, 2005, pp. 396--403.

\bibitem{KschischangFreyLoeliger01}
F.~R. Kschischang, B.~J. Frey, and H.-A. Loeliger, ``Factor graphs and the
  {Sum-Product Algorithm},'' \emph{IEEE Trans. Inform. Theory}, vol.~47, no.~2,
  pp. 498--519, Feb. 2001.

\bibitem{McElieceMacKayCheng98}
R.~J. McEliece, D.~J.~C. MacKay, and J.-F. Cheng, ``Turbo decoding as an
  instance of pearl's `belief propagation' algorithm,'' \emph{IEEE J. Select.
  Areas Commun.}, vol.~16, pp. 140--152, Feb. 1998.

\bibitem{BraunsteinZecchina04}
\BIBentryALTinterwordspacing
A.~Braunstein and R.~Zecchina, ``Survey propagation as local equilibrium
  equations,'' \emph{Journal of Statistical Mechanics: Theory and Experiment},
  vol. 2004, no.~06, p. P06007, 2004. [Online]. Available:
  \url{http://stacks.iop.org/1742-5468/2004/P06007}
\BIBentrySTDinterwordspacing

\bibitem{SunZhengShum03}
J.~Sun, N.-N. Zheng, and H.-Y. Shum, ``Stereo matching using belief
  propagation,'' \emph{IEEE Transactions on Pattern Analysis and Machine
  Intelligence}, vol.~25, no.~7, pp. 787--800, 2003.

\bibitem{Tanaka02}
\BIBentryALTinterwordspacing
K.~Tanaka, ``Statistical-mechanical approach to image processing,''
  \emph{Journal of Physics A: Mathematical and General}, vol.~35, no.~37, pp.
  R81--R150, 2002. [Online]. Available:
  \url{http://stacks.iop.org/0305-4470/35/R81}
\BIBentrySTDinterwordspacing

\bibitem{HeskesAlbersKappen03}
T.~Heskes, C.~Albers, and H.~J. Kappen, ``{A}pproximate {I}nference and
  {C}onstrained {O}ptimization,'' in \emph{Proc. of the 19th Annual Conf. on
  Uncertainty in Artificial Intelligence (UAI-03)}.\hskip 1em plus 0.5em minus
  0.4em\relax San Francisco, CA: Morgan Kaufmann Publishers, 2003, pp.
  313--320.

\bibitem{YedidiaFreemanWeiss05}
J.~S. Yedidia, W.~T. Freeman, and Y.~Weiss, ``Constructing free-energy
  approximations and {Generalized} {Belief} {Propagation} algorithms,''
  \emph{IEEE Transactions on Information Theory}, vol.~51, no.~7, pp.
  2282--2312, July 2005.

\bibitem{Minka01}
T.~Minka, ``{Expectation Propagation} for approximate {Bayesian} inference,''
  in \emph{Proc. of the 17th Annual Conf. on Uncertainty in Artificial
  Intelligence (UAI-01)}.\hskip 1em plus 0.5em minus 0.4em\relax San Francisco,
  CA: Morgan Kaufmann Publishers, 2001, pp. 362--369.

\bibitem{OpperWinther05}
M.~Opper and O.~Winter, ``{Expectation Consistent} approximate inference,''
  \emph{Journal of Machine Learning Research}, vol.~6, pp. 2177--2204, Dec.
  2005.

\bibitem{WeissFreeman01}
Y.~Weiss and W.~T. Freeman, ``On the optimality of solutions of the max-product
  belief-propagation algorithm in arbitrary graphs,'' \emph{IEEE Transactions
  on Information Theory}, vol.~47, no.~2, pp. 736--744, Feb. 2001.

\bibitem{BraunsteinMezardZecchina05}
\BIBentryALTinterwordspacing
A.~Braunstein, M.~M{\'e}zard, and R.~Zecchina, ``Survey propagation: An
  algorithm for satisfiability,'' \emph{Random Structures and Algorithms},
  vol.~27, no.~2, pp. 201--226, 2005. [Online]. Available:
  \url{http://dx.doi.org/10.1002/rsa.20057}
\BIBentrySTDinterwordspacing

\bibitem{WiegerinckHeskes03}
W.~Wiegerinck and T.~Heskes, ``Fractional belief propagation,'' in
  \emph{Advances in Neural Information Processing Systems 15}, S.~T. S.~Becker
  and K.~Obermayer, Eds.\hskip 1em plus 0.5em minus 0.4em\relax Cambridge, MA:
  MIT Press, 2003, pp. 438--445.

\bibitem{TatikondaJordan02}
S.~C. Tatikonda and M.~I. Jordan, ``{Loopy Belief Propogation} and {Gibbs}
  {Measures},'' in \emph{Proc. of the 18th Annual Conf. on Uncertainty in
  Artificial Intelligence (UAI-02)}.\hskip 1em plus 0.5em minus 0.4em\relax San
  Francisco, CA: Morgan Kaufmann Publishers, 2002, pp. 493--500.

\bibitem{Tatikonda03}
S.~C. Tatikonda, ``Convergence of the sum-product algorithm,'' in
  \emph{Proceedings 2003 IEEE Information Theory Workshop}, 2003.

\bibitem{IhlerFisherWillsky04}
A.~T. {Ihler}, J.~W. {Fisher}, and A.~S. {Willsky}, ``Message errors in
  {Belief} {Propagation},'' in \emph{Advances in Neural Information Processing
  Systems 17 (NIPS*2004)}, L.~K. Saul, Y.~Weiss, and L.~Bottou, Eds.\hskip 1em
  plus 0.5em minus 0.4em\relax Cambridge, MA: MIT Press, 2005, pp. 609--616.

\bibitem{IhlerFisherWillsky05}
------, ``Loopy {Belief} {Propagation}: Convergence and effects of message
  errors,'' \emph{Journal of Machine Learning Research}, vol.~6, pp. 905--936,
  2005.

\bibitem{Heskes04}
T.~Heskes, ``On the uniqueness of {Loopy} {Belief} {Propagation} fixed
  points,'' \emph{Neural Computation}, vol.~16, no.~11, pp. 2379--2413, Nov.
  2004.

\bibitem{Weiss00}
Y.~Weiss, ``Correctness of local probability propagation in graphical models
  with loops,'' \emph{Neur. Comp.}, vol.~12, pp. 1--41, 2000.

\bibitem{Dieudonne69}
J.~Dieudonn{\'e}, \emph{Foundations of Modern Analysis}, ser. Pure and Applied
  Mathematics, P.~Smith and S.~Eilenberg, Eds.\hskip 1em plus 0.5em minus
  0.4em\relax New York: Academic Press, 1969, vol. 10-I.

\bibitem{Deutsch75}
E.~Deutsch, ``On matrix norms and logarithmic norms,'' \emph{Numerische
  Mathematik}, vol.~24, no.~1, pp. 49--51, Feb. 1975.

\bibitem{VanDenEssen97}
A.~Cima, A.~van~den Essen, A.~Gasull, E.~Hubbers, and F.~Manosas, ``A
  polynomial counterexample to the {Markus-Yamabe Conjecture},'' \emph{Advances
  in Mathematics}, vol. 131, no.~2, pp. 453--457, Nov. 1997.

\bibitem{Georgii88}
H.-O. Georgii, \emph{Gibbs Measures and Phase Transitions}.\hskip 1em plus
  0.5em minus 0.4em\relax Berlin: Walter de Gruyter, 1988.

\bibitem{TagaMase06}
N.~Taga and S.~Mase, ``On the convergence of loopy belief propagation algorithm
  for different update rules,'' \emph{IEICE Trans. Fundamentals}, vol. E89-A,
  no.~2, pp. 575--582, Feb. 2006.

\bibitem{MooijKappen05c}
\BIBentryALTinterwordspacing
J.~M. Mooij and H.~J. Kappen, ``On the properties of the {Bethe} approximation
  and {L}oopy {B}elief {P}ropagation on binary networks,'' \emph{Journal of
  Statistical Mechanics: Theory and Experiment}, vol. 2005, no.~11, p. P11012,
  2005. [Online]. Available: \url{http://stacks.iop.org/1742-5468/2005/P11012}
\BIBentrySTDinterwordspacing

\bibitem{ElidanMcGrawKoller06}
G.~Elidan, I.~McGraw, and D.~Koller, ``{R}esidual {B}elief {P}ropagation:
  {I}nformed {S}cheduling for {A}synchronous {M}essage {P}assing,'' in
  \emph{Proceedings of the Twenty-second Conference on Uncertainty in AI
  (UAI)}, Boston, Massachussetts, July 2006.

\end{thebibliography}

\end{document}